\documentclass[fleqn,usenatbib,useAMS]{mnras}

\usepackage[T1]{fontenc}
\usepackage{ae,aecompl}
\usepackage{graphicx}	
\usepackage{amsmath}	
\usepackage{amssymb}	
\usepackage[british]{babel}
\usepackage{multicol}        
\usepackage{bm}		
\usepackage{pdflscape}	
\usepackage{times}
\usepackage{float}
\usepackage{caption}
\usepackage{url}

\title[Scalar Super-massive Black Holes]{On the Possibility that Ultra-Light Boson halos host and form Super-massive Black Holes}

 \author[Avilez et al.]
       {Ana A. Avilez$^1$,\thanks{E-mail: aavilez@fis.cinvestav.mx}
       Luis E. Padilla$^1$,\thanks{E-mail: 
       lpadilla@fis.cinvestav.mx}
       Tula Bernal$^2$\thanks{E-mail: 
       tbernal@fis.cinvestav.mx}
       and Tonatiuh Matos$^1$\thanks{E-mail: tmatos@fis.cinvestav.mx}
       \thanks{Part of the Instituto Avanzado de Cosmolog\'ia
	   (IAC) collaboration (http://www.iac.edu.mx)}
       \\
	   $1$ Departamento de F\'{\i}sica, Centro de Investigaci\'on y de
       Estudios Avanzados del IPN, AP
       14-740, Ciudad de M\'exico 07000, M\'exico\\
       $2$ Universidad Aut\'onoma Chapingo, km.~38.5 Carretera
       M\'exico-Texcoco, 56230, Texcoco, Estado de M\'exico, M\'exico
       }

\date{\today}

\pubyear{2016}

\begin{document}
\label{firstpage}
\pagerange{\pageref{firstpage}--\pageref{lastpage}}
\maketitle

\begin{abstract}
Several observations suggest the existence of super-massive black holes (SMBH) at the centers of giant galaxies. 
However the mechanism under which these objects form remains non completely understood.
In this work we review an alternative mechanism of formation of galactic SMBHs. This is, the collapse of   
ultra-light scalar field configurations playing the role of dark matter halos. Several works have studied this scenario and investigated its plausibility. They 
demonstrate, by different techniques that a scalar field configuration with masses larger than $0.6\,m^2_{Pl}/m_\phi$ is able to form a black hole ( configurations made of ultra-light bosons with mass  $m_\phi\sim10^{-22}eV/c^2$ has a critical mass of collapse of $10^{13}M_\odot$). I has been shown that the collapse into a SMBH  occurs at a very slow rate, whilst the remaining forms the galactic halo and it is described by quasi-resonant solutions with cosmic 
lifetime. The theory of collapse of scalar field configurations into compact objects has been extensively studied in a full theoretical way mostly regarding to
boson stars. In this work we extrapolate such results for models of galactic halos hosting central SMBHs. We construct
the simplest setup of an ultra-light scalar field configuration laying in a Schwarzschild space-time for modeling a galactic system with a SMBH in the quasi-static limit.
This model is applicable to systems either out of their accretion-phase at late times or either for those undergoing into a very slow accretion-phase, such as some early-type giant elliptic galaxies and bulges.
 On the other hand, very recent direct observation of Sagittarius A by the Event Horizon Telescope \citet{RicarteDexter2015} will open an era of explorations of the deep-inner galactic region of the Milky Way that will possible to test and distinguish various dark matter models. Thus we use our model to give a very first step towards that direction. We derive the stellar kinematics induced by this sort dark matter to obtain information of the black hole's influence into the observable features of the velocity field of visible matter deep inside the galaxy. Thus we compute the radial velocity, acceleration and velocity dispersion densities for some realistic cases. 
\end{abstract}


\section{Introduction}
\label{introduction}

A host of observations indicate the existence of super-massive black holes (SMBH) placed in the center of most luminous galaxies. SMBHs are enormous in comparison to stellar black holes, their masses may run between $10^6 - 10^{10} M_\odot$ \citet{Lynden1969,Capellari2011,McConnellNature}.
An open problem regarding SMBHs based on observations is why their masses $M_{bh}$, 
at the centers of normal early-type galaxies and bulges correlate with various global properties of the stellar components. The most important relationships occur between $M_{bh}$ and the bulge mass and a tighter correlation between $M_{bh}$ and the stellar velocity dispersion of the host galaxy bulge 
firstly reported by 
\citet{MerrittFerrarese:2001} and 
\citet{Gebhardt:2000}, suggesting that the formation and evolution of SMBHs and the bulge of the parent galaxy may be closely related.  

Although many research groups are trying to understand how these objects were formed their origin is still mysterious due to their huge mass and the large red-shift at which they have been observed \citet{Netzer2003,Genzel2017}. Between the most popular proposals the mechanism of formation of SMBHs goes as follows: likewise stellar black holes which result from the collapse of massive stars, SMBHs are produced by the collapse of massive clouds of gas during the early stages of formation of a galaxy \citet{SilkRees1998}. Another common idea has been that a stellar black hole become super-massive by accretion of large amounts of baryonic material and cold dark matter over long time. Another suggestion is that a cluster of stellar black holes are formed and eventually merge into a SMBH \citet{Menou2001}.

However, so far under this standard scenario there is not a fully satisfactory explanation for the formation and evolution of such SMBHs at such large redshift, even taking into account effects coming from baryonic matter physics. 
For that reason, in this work we consider a novel alternative of formation of SMBH starting from the hypothesis that galactic halos
are made of ultra-light scalar-field configurations playing the role of dark matter all laying in the ground state and therefor making up a Bose-Einstein condensate, which at certain moment in the cosmic history, when it reaches a critical mass, it starts collapsing into a black hole . We argue that the system formed by the black hole and the dark matter halo corresponds to a quasi-static unstable system undergoing into collapse in a time-length as large as the age of the universe \citet{Barranco:2011,Barranco:2012}. Recent detections of SMBH in the centers of dense dwarf galaxies  \citet{AhnSeth2017} strongly support this idea. The fact that the SMBH masses are of the same order of magnitude than their host galaxies suggests that these galaxies were giant galaxies at early times  . An interesting result supporting our hypothesis is that these galaxies contain a dominant fraction of dark matter and hence the accretion rate of the visible fraction of matter is larger than the dark-matter rate, this is in agreement with results from \citet{Barranco:2011,Barranco:2012,Chavanis:2016,UrenaLiddle:2002du} which are the preamble of our hypothesis. 
It is worth to make a parenthesis here in order to stress that this sort of dark matter candidate is a strong and well studied alternative to the cold dark matter (CDM) paradigm. It is usually known as scalar field dark matter (SFDM) model which proposes that dark matter is made of bosonic excitations of an ultra-light scalar field. Typically, these modes are interpreted as spin-0 particles with ultra-light masses of $m$$\sim$$10^{-22} eV/c^2$, which might include self-interactions. One important feature of the SFDM model is that it forms structure in the same way as the CDM \citep{suarez:2011,magana:2012}, reproducing the cosmic microwave background and the mass power spectrum at large scales \citep{RodriguezMontoya:2010,hlozek:2015,schive:2015}. It has been shown that these ultra-light bosons form Bose-Einstein Condensates (BEC) at cosmological scales, making up the galaxy halos \citep{Matos-Urena:2001}. In the cosmological scenario, the ultra-light masses of these particles prevent the over-abundance of small halos, as well as the ``cuspy'' density profiles in galaxies, challenges faced by the CDM model as showed by the current observations.

In this work we consider this last alternative, motivated by the natural solution of this model to the CDM problems. 
The idea of the scalar fields as the dark matter of the universe was first mentioned in \citet{ruffini:1983}; since then the idea was rediscovered several times using different names, for example: Scalar Field Dark Matter (SFDM) \citet{Matos-Guzman:1998}, Fuzzy DM \citet{Hu:2000}, Wave DM \citet{bray:2010,schive:2014}, Bose-Einstein Condensate DM \citet{Bohmer-Harko:2007} or Ultra-light Axion DM \citet{Marsh-Ferreira:2010}  \citet[see also][among others]{membrado:1989,Spergel:1989,sin:1994,Ji:1994,lee:1996,Sahni:1999,Peebles:2000,Goodman:2000,Matos-Urena:2000,
Matos-Urena:2001,Wetterich:2001,Arbey:2001a,Woo:2009,Lundgren:2010,Calabrese:2016,ostriker:2016}.
However, the first systematic study of this idea started in \citet{matos:1998,matos:1999}, showing that the rotation curves in the model fit very well with the observations of galaxies.
The cosmology was studied for the first time in \citet{matos:2000b,Matos-Urena:2001}, where using the amount of satellite galaxies we observe in the Milky Way vicinity, 
Since in this work we are interested in collapsing scalar field configurations, let us now briefly review the history of this model within the context of bound scalar field systems and their stability conditions. For an ultra-light scalar field model with mass $m\sim 10^{-22}eV/c^2$  \citet{Alcubierre:2002a} found that the gravitational collapse of scalar field configurations forms quasi-stable objects with masses of the order of a halo of a galaxy ($\sim10^{12} M_\odot$). Other systematic studies of the SFDM were performed by \citet{Arbey:2001a,Arbey:2001b} and more recently by \citet{Marsh-Ferreira:2010,Marsh:2015b,schive:2014}.

	In a series of papers, \citet{matos:2007,bernal:2008,robles:2012} showed that the SFDM configurations are able to model core halos \citep[see also][]{Harko:2011}. The gravitational lensing produced by these objects has been explored in \citet{nunez:2010,Robles:2013}. A self-interacting scalar field with a $\phi^4$ potential has been studied in \citet{matos:2011,Abril-Chavanis2016}. Numerical simulations of galaxies showing the formation of the characteristic spiral arms and bars were performed in \citet{medina:2014} and \citet{bray:2015}. \citet{robles:2015} obtained, with numerical simulations, that satellite galaxies are stable around SFDM halos.
The SFDM at finite temperature is an interesting result given in 
\citet{Robles:2013}.


As mentioned above, here we start from the hypothesis that galactic halos are made of SFDM configurations whose mass reached the critical value at some point early in the history of the universe.
The problem of collapse of a scalar field configuration into a black hole has been extensively studied during the last few decades in many contexts. Firstly, by analyzing different regimes of stability (or instability) of scalar field solutions, different groups have arrived to remarkable insights about self-gravitating BECs \citep{Seidel-Suen:1990,Seidel-Suen:1991,Seidel-Suen:1994,SanchisGual:2016,Alcubierre:2002a,Alcubierre:2002b,CruzGuzman2011, Barranco:2011,Barranco:2012,Chavanis:2016}. The typical systems considered for these studies have been boson-stars and therefore they are limited to explain the formation of stellar black holes. In this work we aim to extend their approaches for large galaxies where SMBHs are typically observed.
The formation mechanism of SMBHs suggested by all these works, is able to explain the origin even of the biggest SMBH observed so far with mass $M$$\sim$$10^{10} M_\odot$. According to some pioneer numerical results of \citet{Seidel-Suen:1990,Seidel-Suen:1994,Alcubierre:2002a}, a SFDM halo with mass above $10^{13} M_\odot$ would undergo into gravitational collapse ending up into a black hole at the center of the system. On the other hand, numerical simulations from \citet{Barranco:2011,Barranco:2012,SanchisGual:2016} show that only one part of the scalar field collapses into a BH and the rest of the field remains around surrounding the BH for a very long time. Although the no-hair theorems prevent static solutions (hair) to exist around the black hole \citep{BekensteinNoHair-1995,NoHair-2015}, nothing prevents these quasi-resonant solutions (wigs) to exist around the BH and they can clearly play the role of the dark matter halo of the galaxy \citep{Myung2008}. In short words the main two hypotheses of this work read as:  i) the formation of super-massive black holes (SMBH) is due to the collapse of a dark matter halo as a scalar field configuration and ii) in the nowadays galactic systems the dark matter halo corresponds to a quasi-resonant solution of the scalar field decaying very slowly into the SMBH, such that it lasts a cosmic-time-scale lifetime.
The self-gravitating system forming a galaxy, under the unstable regime, is described by a very cumbersome and highly non-linear set of coupled equations. To handle both the physics of the black hole ($10^{-6} \ pc$) and that of the galaxy ($10-30 \ kpc$) by numerical means, is a tremendous task. However, previous numerical work regarding stellar systems can be very useful to give a very first step in that direction. With this work we aim starting to address this problem in a semi-analytical way, departing from the very simplest scenario: we assume that the galactic system is formed by a SFDM halo laying in a Schwarzschild space-time. This is valid if we consider the quasi-static limit, in this case we are assuming that the black hole has been already formed and that the remaining SFDM halo corresponds to a long-living quasi-resonant solution. We solve analytically the Klein-Gordon equation in a Schwarzschild space-time (SKGE) with spherical symmetry assuming that the observer is placed far away from the black hole and inside the galactic bulge. Other SFDM halos have been found as the gravitationally bound solutions of the Schroedinger-Poisson and Gross-Pitaevskii-Poisson systems, which are the weak-gravity version of the Klein-Gordon and Einstein equations in the Newtonian limit, neglecting or including self-interactions. It has been shown that such an approach suffices to reproduce the galactic dynamics in the Newtonian regime, e.g. the stellar rotation curves of many galaxies \citep{Harko:2011,Robles:2013,medina:2014,Bernal:2017}, the velocity dispersion in dwarf spheroidal galaxies \citet{Martinez-Medina:2015,Chen-Schive:2016,Gonzalez-Marsh:2016} and the dynamical masses inferred from X-ray galaxy clusters observations \citet{Bernal:2016}, among others.

In the present work, we have found that our analytical solution is indistinguishable from Schroedinger-Poisson solutions in the  outer region of galaxies where rotation curves are measured. The dark matter density profile is only modified close to the center, inside the bulge, where SMBHs are detected from observations of the kinematics of visible matter. We refer to these kind of solutions as Schwarzschild-scalar-field-dark-matter (SSFDM). In despite that such region is yet far away from the SMBH, it is amazing that the mass of the SMBH and the dispersion of velocities of stars and gas are correlated  \citet{MerrittFerrarese:2001,Gebhardt:2000,Ferrarese:2004,McConnell:2012}. Usually this connection is attributed to accretion winds feedback of the black hole into the baryonic matter laying in the bulge \citet{Larkin:2016}. However, although the accretion winds could highly affect the stellar kinematics up to some point, we argue that given the difference of scales at which both phenomena occur, the correlation between both thresholds still remains as a mystery and may be triggered by a different mechanism.    
As mentioned above, SMBHs are not observed directly, in turn their properties (mass and angular momentum) are inferred from features of the velocity field of stars in the galactic bulge, particularly the velocity dispersion of stars \citet{McConnell:2012}. Therefore, any framework designed to describe the joint dynamics of a SMBH and its
host galaxy should be able to fit, in a reasonable way, the observed stellar velocity dispersion.
In this work, for parameters of real galaxies, we compute theoretical predictions of the mean velocity dispersions corresponding to each solution for a fixed mass of the black hole, for different values for the wave-length 
of the profile. We achieve this numerically by solving the Jeans equation and modeling the spherical density and mass distributions using the phenomenological Plummer profile \citep{Plummer:1911}, which fits well the density distribution describing stellar systems like bulges,
where normally SMBHs are hosted. In order to fully construct a layout of the SFDM profile that actually is compatible with observations, we determine the unfixed central density and characteristic wave-length by using theoretical and observational constrains. Firstly, based on theoretical grounds, we derive an upper bound for the wave-length of the solution using  conditions over the spectrum of solutions proposed in \citet{Barranco:2011}. Secondly, we determine the central density of the profile by assuming the universality of the maximum acceleration of DM particles recently derived from observations of a large set of galaxies \citet{Urena-Robles-Matos:2017}. We found in this case that the dark matter central density is an increasing function of the mass of the black hole, that is consistent with previous results  
\citet{lee:2015,Barranco:2011,Barranco:2012}.
A 
complementary result of this work is a modification of the constraint over the central surface density given by $\mu_{DM}$$=$$constant$ \citep{Urena-Robles-Matos:2017}. We carry out their analysis for galaxies hosting SMBHs. Finally, we constrain the wave-lengths of the SFDM profiles for different black-hole masses, by fitting mean observed values of the velocity dispersion of a number of elliptical galaxies and bulges. We obtained wave-lengths slightly smaller than those from previous fittings using Wave-DM profiles, however, the presence of the black hole contracts the size of the halo; we demonstrate this by carrying out a perturbative analysis on Wave-DM profiles affected by a central black-hole.   

We would like to point out that, in despite of the simplicity of this model, it certainly has predictive-power since it is able to reproduce the mean values of observables and it is a sufficient reason to keep studying it and improving it. Moreover, it supports the initial hypothesis of formation of SMBHs from the collapse of SFDM halos. Furthermore, recently an international team of astronomers used a earth-sized telescope formed as a series of telescopes placed across the globe, known as the Event Horizon Telescope (EHT) to obtain for the very first time a direct image of Sagittarius A (the SMBH hosted in the center of the Milky Way) \citet{RicarteDexter2015,EventHorizonTelescope}. Observations of the deep-inner galactic region are very likely to be improved in short time and this would bring up a new source of evidence of the properties of SMBH and its influence on stars laying in the galactic bulge. Particularly, it might be possible to obtain a direct measurements of the mass and angular momentum of these objects. In addition, from observations of the stellar evolution across this region, wealthy information of the dark matter configurations would be inferred which would be useful to discriminate between different DM models.  Either to test our hypothesis about SMBH formation and to compare different DM models in the deep-inner galactic regions using these and upcoming direct observations of SMBH in a short future is a compelling goal that we are after.


The paper is organized as follows: In section~\ref{sec:review} we make a brief review on the research work about the collapse of a scalar-field configuration into black-hole. In section~\ref{sec:wigs} we state our hypotheses. In section~\ref{sec-model}, we present our model and analyse some conditions over the spectrum of solution; we derive 
the general Klein-Gordon equation in a Schwarzschild space-time and then consider the limit for radius far away from the horizon of the SMBH; we dub such equation as SKGE and solve it analytically. Section~\ref{sec:driving} we treat the same equation perturbativelly such that it turns into an equation for a forced oscillator,  doing do helps us to understand the perturbative effect of the BH over Wave-DM solutions in a more intuitive way. In section~\ref{sec:profiles} we compute and study dark matter density and mass profiles from 
our exact solution. In section~\ref{sec:constraints} we constrain the central density parameter of the profile and extend the constraint $\mu_{DM}= constant$ \citep{Urena-Robles-Matos:2017}
for the SSFDM model proposed here. In section~\ref{sec:visible-matter} we constrain the wave-length by fitting measurements of the mean velocity dispersion of a number of real elliptical galaxies hosting SMBHs. Finally, in section \ref{sec:discussion} we 
present our conclusions.

\section{Brief Historical Review of the Problem: Collapse of SFDM Configurations}
\label{sec:review}

The pioneering studies of equilibrium configurations of
self-gravitating bosons in the ground state
\citep{Ruffini-Bonazzola:1969}, show
that the maximum mass of a scalar field configuration  depends on the mass of the boson. This critical mass is understood  as follows: for a system of N self-gravitating bosons such that  $N > N_{crit}$, a critical number of bosons, the gravitational collapse actually occurs.
In the non-self-interacting case they found that the critical mass of collapse is
\begin{equation}
  M_{crit} \sim \frac{m_{Pl}^2}{m} ,
\label{Mcrit}
\end{equation}
where $m_{Pl}$ is the Planck mass and $m$ is the mass of the scalar field. At this critical value, the binding energy per particle becomes comparable to the rest energy. For a system with a number of particles greater than $N_{crit}$ there exists non-equilibrium configurations of self-gravitating bosons in their ground state.
For the ultra-light bosons we study here with masses between $m$$\sim$$10^{-23}-10^{-22} eV/c^2$, the critical
mass of collapse lays between $M_{crit} \sim 2.8 \times 10^{11}-10^{12} M_\odot$. Later, \citet{Seidel-Suen:1991,Seidel-Suen:1994} used numerical spherically symmetric simulations and found that for a mass $m$ of the SF, the critical mass of collapse of a set of bosonic particles is
\begin{equation}
	M_{crit} \sim 0.6 \frac{m_{Pl}^2}{m}.
\end{equation}
%


Later on, \citet{Alcubierre:2002a,Alcubierre:2002b} found through numerical simulations that the critical mass of collapse to form stable solutions from an initial SF perturbation with a $\cosh$-like potential, for a boson mass $m = 1.1 \times 10^{-23} eV/c^2$, is given by $M_{crit} \sim 10^{13} M_\odot$.
%
For the boson mass $m$$\sim$$10^{-22} eV/c^2$, the critical mass of stability is $M_{crit} \sim 10^{12} M_\odot$, the typical mass of a galaxy, and the galactic halos could be born very early in the universe.

For SFDM configurations including excited
states \citep{Seidel-Suen:1990,Hawley:2003,Urena:2009,Bernal:2010,
Urena-Bernal:2010}, the resulting configurations can have larger masses until they reach new equilibrium 
configurations through mass-loss (gravitational cooling).



In the case of a self-interacting scalar field $\phi$ with the simplest $\phi^4$-potential
\begin{equation}
	V(|\phi|^2) = m^2 \phi^2 + \frac{\lambda}{2} |\phi|^4 ,
\end{equation}
the critical mass of collapse is given by \citep{Colpi:1986}
\begin{equation}
	M_{crit} \sim 0.22 \sqrt{\Lambda} ~ \frac{m_{Pl}}{m^2} ,
\end{equation}
where $\Lambda \equiv \lambda/(4\pi G)$, for the self-interaction $\lambda$ and $G$ the gravitational constant. Notice that the self-interaction increases the critical mass for large values of the mass of the SF. In the context of standard axions, their masses run between $10^{-6}-10^{-3}$ eV in order to account for the dark matter. The original axion had masses of $keV-MeV$ and relevant interaction with standard matter and radiation, which according to quantum field theory are easily observable in particle accelerators, however they have been already ruled out. The `invisible axion' runs between $10^{-33}-10^{-3}$ eV. The today so-called `standard axions' have masses between $10^{-6}-10^{-3}$ eV \citet{Axions-review:2010}. Therefore, the self-interaction is crucial in order to make these models observationally relevant. For ultra-light scalar fields, even the free case is observationally appealing since the critical mass is close, in order of magnitude, to that of observed SMBHs; however, the self-interaction in this case plays an important role. On one hand it has been shown that such self-interaction is required in order to these configurations to condense into Bose-Einstein condensates. On the other hand, the self-interacting field offers a richer phenomenology which makes the model more flexible and capable to predict more realistic systems. Recently, \citet{Chavanis:2016} studied analytically, in the Newtonian approximation, the collapse of self-interacting BECs with attractive self-interaction in the Thomas-Fermi limit (dominant self-interactions regime), and found the critical mass and time of collapse of a BEC to form a black hole. In the context of the `standard' axion mass ($m_a$$\sim$$10^{-4} eV/c^2$), in this last work the formation of mini black holes with masses $M$$\sim$$10^{-13} M_\odot$ is predicted.
For ultra-light axions ($m$$\sim$$10^{-20} eV/c^2$), it is predicted  that the black holes with masses of $M$$\sim$$10^5 M_\odot$  are formed (the BH at the center of the Milky Way is $\sim$$10^6 M_\odot$), with reasonable times of collapse of $t$ $\sim$ Myrs.

In what is concerning to this work, our humble model only considers the free field which is enough for our purposes.  Nevertheless, we aim to extend our approach in further works.

\section{Black holes wigs as long-lasting dark matter halos}
\label{sec:wigs}

 Besides the problem of formation of a SMBH through the collapse of SFDM configurations, a further question remains: Once the SMBH is formed, does a self-gravitating and stable scalar galactic halo remain afterwards?
Strictly speaking, no-hair theorems lead us to give a negative answer to the question. However, for realistic circumstances and practical purposes, the answer is surprisingly affirmative.
No-hair theorems condemn these solutions to pass away after some time \citep{BekensteinNoHair-1995,NoHair-2015}; while that is true, it also has been demonstrated that some solutions survive at least a cosmic time. That is, although black holes are condemn always to be bald, nothing prevent black holes to wear wigs along their whole life. The boson excitations making up the wigs are subject to gravity and also to their own dispersive nature inherit by the KG equation. Although a perfect balance between these competing effects can not be achieved, at least the decay time can be controlled choosing a proper mass for the configuration.    

\citet{Barranco:2011,Barranco:2012}, and more recently \citet{SanchisGual:2016}, studied analytically and numerically configurations of scalar field embedded in a Schwarzschild space-time, once the BH has been formed. They realized that it is possible to find physically meaningful, long-lived SF configurations. In particular, for ultra-light scalar fields laying around SMBHs and axions around primordial BHs. They found that for masses ($m < 10^{-22} eV/c^2$) and SMBHs of $M < 5 \times 10^{10} M_\odot$, the configuration can survive for times larger than $10^{10}$ yrs. In this sense, their results strongly support the hypothesis that the dark matter is a scalar field in the galaxies hosting SMBHs. Furthermore, the whole dynamics of the system, including its formation and evolution along the cosmic history, arises from a single physical framework without aid of baryonic physics.

\section{THE MODEL: ULTRA-LIGHT SCALAR FIELD CONFIGURATIONS IN A SCHWARZSCHILD SPACE-TIME}
\label{sec-model}

We start studying the simplest model for a SFDM halo hosting a black hole.  
We assume that the geometry of the space-time surrounding a black hole is described by the Schwarzschild metric which, in spherical coordinates, is given by
\begin{eqnarray}
 ds^2 = -\left( 1 - \frac{2M}{r} \right)  ~ d t^2 + \left( 1 - \frac{2M}{r} \right) ^{-1} d r^2 + r^2 d^2\Omega,
 \label{eq:Schwmetric}
\end{eqnarray}
where $M$ is the mass of the black hole in units of distance (from hereafter we use units $G=c=1$) and $d^2\Omega\equiv d \theta^2 + \sin^2 \theta ~ d \varphi^2$ is the solid angle square differential.

The dynamics of the scalar field described by the Klein-Gordon equation written in a Schwarzschild background space-time is given by 
\begin{eqnarray}
  -\partial_t^2 \phi + \frac{g}{r^2}\partial_r\left[r^2 g~\partial_r \phi \right] - 
  g\frac{L_{\theta,\varphi}(\phi)}{r^2}-g\left(m^2+\lambda \phi^2\right)\phi=0 ,
\label{eq:SchwKG}
\end{eqnarray}
where we have assumed $r \neq 2M$ to avoid singular points and we have introduced the following definitions of the angular-momentum operator and the $g$ function:
\begin{eqnarray}
  L_{\theta,\varphi}(\phi) \equiv \frac{1}{\sin\theta} \left( \sin \theta ~ \partial_{\theta} 
  \phi \right) + \frac{1}{\sin^2\theta} \partial_\varphi^2 \phi ;
\end{eqnarray}
\begin{eqnarray}
  g(r) \equiv \left( 1 - \frac{2M}{r} \right) .
\label{eq:g}
\end{eqnarray}
 
It is easy to realize that equation~\eqref{eq:SchwKG} is a separable equation with respect to time and space coordinates. Thus, it admits solutions of the form
\begin{eqnarray}
  \phi(t,r,\theta,\varphi) = \psi(r,\theta,\varphi) ~ e^{-i \omega t} ,
\label{eq:SeparationS-T}
\end{eqnarray}
with harmonic time dependence for an arbitrary frequency $\omega$.
After plugging equation~\eqref{eq:SeparationS-T} into~\eqref{eq:SchwKG} we obtain
\begin{eqnarray}
  \left( \omega^2 - g m^2 \right) \psi + \frac{g}{r^2} \partial_r \left[ r^2  g 
  ~ \partial_r \psi \right] - g \frac{L_{\theta,\varphi}(\psi)}{r^2} - g \lambda
  \psi^3 = 0.
\end{eqnarray}
By the moment, in this work we set $\lambda=0$ for sake of simplicity, in which case the equation is linear and admits solutions of the form 
\begin{eqnarray}
  \psi(r,\theta,\varphi) = R_l(r) ~ Y_l^n(\theta,\varphi) ,
\end{eqnarray}
where $R_l$ is a function depending only on the radial coordinate $r$ and the angular solution is given by the spherical harmonics $Y_l^n$ for non-negative integers $l \geq |n|$ 
After using equation~\eqref{eq:g}, the radial equation is given by
%
\begin{eqnarray}\label{eq:KGradial} 
  m^2 R_l - \frac{g}{r^2} \partial_r \left[ r^2 g ~ \partial_r R_l \right] +
  g \frac{l(l+1)}{r^2} - \frac{2Mm^2}{r} R_l = \omega_l^2 R_l .
\end{eqnarray}
Notice that the last term on the LHS in equation~\eqref{eq:KGradial} becomes small far away from the BH's event horizon. In the case $2Mm^2/r\rightarrow 0$, equation~\eqref{eq:KGradial} turns into the free Schr\"odinger equation. This last point is rather important in what follows. The length-scale $2Mm^2$ shall turn out to be a natural measure of the size of the configuration. Even though the scale of the 
black hole is very different to that at which the galactic dynamics occur, masses $M$ for SMBHs give rise to $2Mm^2\sim kpc$. Moreover, since we are interested in describing the phenomenology occurring inside the galactic bulge close to its edge, that is, at scales of few $kpc$, solutions  of~\eqref{eq:KGradial} in the regime where $r>2Mm^2$ are actually what we are looking for. In the following sections we are going to handle equation~\eqref{eq:KGradial} in such a limit. Before doing so, 
 we shall stop to analyse the spectrum of solutions arising for the range of parameters relevant for SMBHs.

\subsection{ABOUT THE EIGENVALUE PROBLEM FOR THE KLEIN-GORDON-SCHWARZSCHILD SYSTEM}
\label{sec-spectrum}

For given masses of the scalar field and the black hole, the spectrum of solutions has been determined numerically and semi-analytically in \citet{Barranco:2011,Barranco:2012}. In that work, they follow a standard procedure: they use convenient radial coordinates so that the KG equation can be transformed into a Schroedinger-like equation with a corresponding effective potential. They demonstrate that the frequencies of the solution cannot be larger than the depth of the well-potential (which is dubbed as ``resonance band'').  Besides, this is equivalent to the standard procedure of solving the Schroedinger equation analytically and to use boundary conditions to determine the full spectrum allowed at each physical setup. A similar analysis can be done by analyzing the KG equation with parameters in a range of values corresponding to models for galactic halos.
Here we follow such a procedure, including realistic values of the parameter $\alpha \equiv Mm$, corresponding to realistic masses of SMBHs which run between $10^{6}-10^{10} M_{\odot}$, assuming a mass of the scalar field $m=10^{-22}~eV$ (see Table~\ref{tb:spectrum}) ~\citep{McConnell:2012,Larkin:2016}. 

For convenience, we can pick coordinates such that equation~\eqref{eq:KGradial} takes a fully
hyperbolic form; in the radial case we use the Regge-Wheeler tortoise coordinate defined as
$r^* \equiv r+2M\ln(r/2M-1)$, such that equation~\eqref{eq:KGradial} can be written as 
\begin{eqnarray}
  - \partial_{r^*}^2 Q^l + V_{eff} Q^l = \omega_l^2 Q^l ,
\label{eq:eigen}
\end{eqnarray}
where $Q^l\equiv r R^l$, and we have introduced an effective potential given by
\begin{eqnarray}
  V_{eff}(r;l,m,r_s) \equiv g(r) \left[ m^2 + \frac{l(l+1)}{r^2} + \frac{2M}
  {r^3} \right] .
\label{eq:Veff}
\end{eqnarray}
By solving exactly the eigenvalue problem established by equation~\eqref{eq:eigen}, we would be able to determine the full spectrum of solutions allowed for this setup. However, in real galaxies, the SFDM halo coexists with stars and gas which might modify considerably the features of the system. For that reason and for simplicity we consider it is reasonable to set the wave-length of the solution $r_s\equiv1/k$ (with $k^2\equiv\omega_l^2-m^2$) as a free parameter and to determine an above cut-off  $k_{max}$ (see Table~\ref{tb:spectrum}).
Let us now consider the spherically symmetric case with $l=0$. For $m$$\sim$$ 10^{-22}~eV$, the potential wells for realistic cases are quite shallow (as shown in Table~\ref{tb:spectrum} and Fig.~\ref{fig1:Veff}). As a consequence, even for the most massive
black holes observed so far, the resonance band is pretty narrow. This suggests that the spectrum of solutions is almost empty for the lightest BH and hence such solutions can be approximated as a single state with $\omega\sim m$, as assumed \citep[see also][]{UrenaLiddle:2002du}.

\begin{minipage}{0.4\textwidth}
\vspace{3mm}
\centering
  \includegraphics[width=\textwidth]{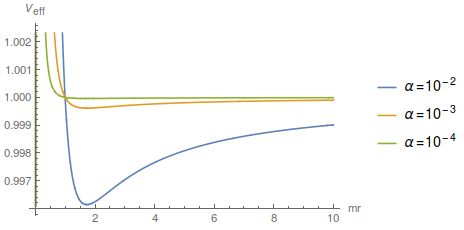}
\captionof{figure}{\small{The effective potential for the general Schwarzschild-Klein-Gordon equation for different values of the parameter $\alpha$. For  configurations to form dark 
matter halos hosting SMBHs, smaller values should be considered, and pretty shallow potentials arise. As a consequence, very large wave-lengths are expected which is in agreement with other values of $r_s$ obtained from fits to rotation curves of galaxies~\citep{Bernal:2017}.}}
\label{fig1:Veff}
\end{minipage}

\begin{minipage}{0.4\textwidth}
\vspace{3mm}
\centering
\begin{footnotesize}
\centering
\begin{tabular}{cccc}
\hline
$M/M_\odot$ & $\alpha$ & $k_{max}/m$ & $r_s^{min}/kpc$ \\
\hline \hline
$10^{8}$&$10^{-5}$  &  $2\times 10^{-3}$ & $0.25$\\
$10^{9}$&$10^{-4}$  &  $6\times 10^{-3}$ & $0.083$\\
$10^{10}$&$10^{-3}$  &  $2\times 10^{-2}$ & $0.025$\\
$1.2\times10^{10}$&$2.5\times10^{-3}$  &  $3\times 10^{-2}$ & $0.017$\\
$2.5\times 10^{10}$&$5\times10^{-3}$  &  $4\times 10^{-2}$ & $0.013$\\
$5\times 10^{10}$ & $7.5\times 10^{-3}$  &  $5\times 10^{-2}$ & $0.010$\\
\hline
\end{tabular}
\end{footnotesize}
\captionof{table}{\small{Maximum wave number and minimum characteristic wavelength of a single state solution of the scalar field for a range of parameters including typical observational masses of SMBHs. Quasi-bound states of the system have square frequencies laying in the resonance band $(V_{eff}(x_{min}),m^2)$.}}
\label{tb:spectrum}
\vspace{3mm}
\end{minipage}


According to \citet{Barranco:2011,Barranco:2012},
stationary modes with real frequency $\omega^2$$>$$m^2$ do not exist. They found that the spectrum of solutions is continuous. The previous statement is in agreement with no-hair theorems. Furthermore these modes do not decay at spatial infinity. 
If $0$$<$$\omega^2$$<$$m^2$, then the modes have purely imaginary frequencies and they form a discrete set for
which the amplitude inside the potential-well takes very
large values when compared with the amplitude close to
the horizon; for that reason they are called \textit{stationary resonances} within the band $V_{eff}$$<$$\omega^2$$<$$m^2$. When the conditions of no waves coming from the horizon and the requirement of spatial infinity decay are imposed, the spectrum of stationary resonances become discrete and complex. This set of solutions has been called \textit{quasi-resonances} in the literature \citep{OhashiSakagami:2004}.
Unfortunately, both sorts of solutions are non-physical due to the
conserved energy density corresponding to a Killing vector, which diverges at the horizon. 
Nevertheless, by numerical calculations, \citet{Barranco:2011} found healthier solutions dubbed as \textit{dynamical resonances} with finite energy density in all regions. These solutions are damped oscillations of the BH for very long times. The spectrum of such set of solutions is the same than the stationary resonances.
Actually, \citet{Barranco:2012} showed
that the real part of the frequency of the quasi-resonant modes coincides with the frequency of oscillation of the stationary and dynamical resonances, and the imaginary part coincides with the decay rate of the dynamical resonances.
  
Firstly, because a real BH is
surrounded by quasi-resonances that are as long lived as the universe, we can assume by now that they are stationary and then we can, as a first approximation, neglect the
decaying part of the solution controlled by the imaginary part of the frequencies.  

In principle $k$ could be taken as a free parameter (as it is usually done in the 
literature), however we argue that the range of its values is restricted by $\alpha$. As mentioned above, $k$ is allowed to take values below $k_{max}$.
However, because $\alpha$ takes such small values for typical galactic SMBH masses, the corresponding effective potentials are quite shallow and 
consequently $k$ can be well estimated by $k_{max}$.

 Usually, within the models accounting for SFDM configurations as galactic halos, the values for $k$ run among $0.35 ~kpc$ and $10 ~kpc$ \citet{Robles-Medina:2015}.
As we shall see below, 
the presence of a black hole produces a considerable reduction in the size of the halo, therefore the estimate we mention is fairly reasonable. For now, let us keep this
assumption, and leave the analysis on whether $k$ can be fully determined from theory or not for a further work.

\subsection{THE SFDM COFIGURATION FAR AWAY FROM THE BLACK HOLE}
\label{sec-SFDMsol}

Because in the following we aim to study the behavior of visible matter at regions far away from the black hole, we need to solve equation~\eqref{eq:KGradial} in the limit in which $2M/r\rightarrow 0$, that is 
\begin{eqnarray}
  k^2 R_l  + \frac{1}{r^2} \partial_r \left[ r^2 ~ \partial_r R_l \right] - \frac{l(l+1)}{r^2} + \frac{2Mm^2}{r}R_l = 0.
\label{eq:KGradialfaraway} 
\end{eqnarray} 
 This equation is valid  for most of the galaxy bulge region.
 The evolution of the SMBH and the galactic system lay in very different spatial scales. Even stars living in the deepest galactic regions well inside
the bulge are parsecs away from the center, while the Schwarzschild radius of the SMBH is one part in a million smaller. On one hand, $2M\sim 10^{-14}~pc$ for the sun and $2M\sim 10^{-6}~pc$ for a SMBH with $M\sim 10^{10} M_\odot$. On the other hand, typically the size of galactic bulges runs from $1-10 ~kpc$.

By taking again the change of variable $Q_l \equiv r~R_l$, equation~\eqref{eq:KGradialfaraway} becomes
\begin{eqnarray}
 \partial_{rr} Q_l + k^2 Q_l - \frac{l(l+1)}{r} + \frac{2Mm^2}{r} Q_l = 0 .
\end{eqnarray}
%
%
For $l=0$, which is sufficient for the spherically symmetric case, a solution is 
\begin{eqnarray}
  \psi(r) = \psi_0 ~ e^{-ikr} ~ {}_1F_1 \left(1-i\frac{Mm^2}{k},2,i2kr \right).
\label{eq:KGSolutionl0}
\end{eqnarray}
where $\psi_0$ is a constant and ${}_1F_1$ is the hyper-geometric function of order $(1,1)$. From now on 
we shall refer to linear combinations of this sort of solutions of as SSFDM configurations.
We shall analyse the features of the profiles arisen from this solution in detail in section~\ref{sec:profiles}.
Well inside the galactic bulge it happens that $kr \ll 1$, therefore the solution approximates to 
\begin{eqnarray}
  \psi \simeq \psi_0~(1-Mm^2r) + \mathcal{O}[(kr)^2]+ \mathcal{O}[(Mm^2)^2] .
\end{eqnarray}
This profile is in agreement to that proposed in \citet{lee:2015}. For small radii inside the bulge, still far away from the black hole ($ 2M \ll r \ll 1/(Mm^2)$), the profile remains constant. Still inside the halo but closer to the edge of the galactic core ($r\sim r_{max}= 1/(Mm^2) $) the profiles start decaying. 
 In the next section we will show that the black hole produces a driving effect on the halo SF solution and the more massive it is more cuspy is the profile of the halo.

\section{THE DRIVING EFFECT OF THE BLACK HOLE ON THE HALO SOLUTION}
\label{sec:driving}

Although equation~\eqref{eq:KGradial} can be solved exactly and analytically, in this section we treat the problem perturbatively in order to have a more intuitive understanding of the effect that a black hole exerts onto a bare SFDM solution. From equation~\eqref{eq:KGradial} 
it can be noticed that the last term is actually a perturbation to the KG equation in flat space for a region $r\gg2M$. Thus we can split the scalar field into a bare solution plus a small perturbation induced by the black hole:
\begin{eqnarray}\label{eq:psipert}
\psi = \bar{\psi} + \delta \psi + \mathcal{O}(M^2) ,
\end{eqnarray}
being $\bar{\psi}$ the solution for the halo in flat space-time given by 
\citep{Robles:2013}
%
\begin{eqnarray}
  \bar{\psi}(r)^k=\bar\psi_0 \frac{\sin(kr)}{r} ,
\label{eq:densidad}
\end{eqnarray}
which is an exact solution of the KG equation for a SFDM perturbation, that corresponds to the galactic DM halo in the Newtonian limit found in the case of a temperature-corrected SF potential for $T\simeq 0$.  The density profile corresponding to such solution has been proved to be successful in fitting the rotation curves of galaxies \citep{Robles:2013,Bernal:2017}, the velocity dispersions observed in dwarf spheroidal galaxies \citep{Martinez-Medina:2015}, the strong gravitational lensing \citep{Robles-lensing:2013} and the dynamical masses from X-ray observations of galaxy clusters \citep{Bernal:2016}.

By plugging~\eqref{eq:psipert} into~\eqref{eq:KGradialfaraway} and setting $l=0$, we obtain an ordinary differential equation for the perturbation given by 
\begin{eqnarray}\label{eq:O1KGradial}
    \delta R_{,rr} + k^2\delta R = - \bar{\psi}_0~\frac{r_0}{r} ~\frac{\sin\left[(k+dk)r\right]}{r} .
\end{eqnarray}
The previous equation corresponds to a driven harmonic oscillator. Notice that the black hole mass quantifies the amplitude of the external force which has the same functional form than the bare solution and has a wave-number $k+dk$; $dk$ quantifies the closeness to the natural frequency of the oscillator: when $dk\rightarrow 0$, the corresponding solution is in resonance with the external force and its amplitude is enhanced. Equation~\eqref{eq:O1KGradial} holds the following analytic solution:
\begin{eqnarray}\label{eq:persol}
\psi_1^k = \bar\psi_1^k\left(\frac{m}{k}\right)^2\left(\frac{r_s}{r}\right)\left(Ci^{(-)}\cos kr -Si^{(+)}\sin kr\right) ,
\end{eqnarray}
where
\begin{eqnarray}\nonumber
Ci^{(-)} &\equiv& Ci[dk~r] - Ci[(k+dk)~r],\\
\nonumber Si^{(+)} &\equiv& Si[dk~r] + Si[(k+dk)~r],
\end{eqnarray}
are the cosine and sine integral functions.

\begin{minipage}{0.43\textwidth}
\vspace{5mm}
\centering
 \includegraphics[width=\textwidth]{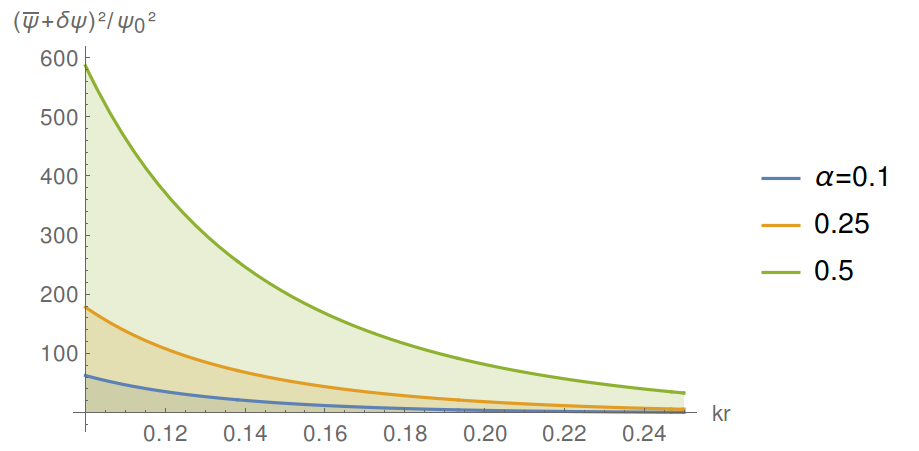}
\captionof{figure}{\small{Dimensionless solutions of the scalar field perturbed by a black hole for different values of the $\alpha$ parameter. As the mass of the black hole increases, the solution becomes more cuspy and its width decreases. This effect is produced by a driving force due to the presence of the black hole.}}
\label{fig:Q0_l0_k1}
\vspace{5mm}
\end{minipage}
\newline

From the behavior of the $Ci^{(-)}$ and $Si^{(+)}$ functions, some physical information about the resonant solution can be extracted. Firstly, the bigger the mass of the BH is, the solution for the scalar becomes more cuspy (see Fig.~\ref{fig:Q0_l0_k1}). As expected, when the frequency of the oscillator is equal to that of the driving force, the amplitude of the oscillator blows-up. This happens because we are considering an idealized situation in which the driving force is formed by a single state of the bare solution; however, a more realistic configuration would correspond to a driven force as a coherent package made of multiple bare solutions with different frequencies laying within frequency band of width $dk$. In such case, $dk$ would never vanish and therefore the solution for the oscillator would never be purely resonant; in turn, while the driving force becomes narrower, the amplitude increases without blowing up. Nonetheless, for the purposes of this subsection, the idealized situation is enough to realize that the perturbative term due to the presence of a black hole works as a driving force which produces enhancement of the amplitude of the solution of the scalar field. Fig.~\ref{fig:Q0_l0_k1} shows solutions in this simple case with $dk\ll1$ for different masses of the black hole.

\section{Dark Matter Mass/Density Profiles}
\label{sec:profiles}

As mentioned above, here we model the system of a SFDM halo and a SMBH assuming that the former is a configuration of a complex scalar field 
laying in a Schwarzschild space-time. This situation corresponds to the static case in which the BH is already formed and stands still 
without accreting any matter or gas and the velocity field is affected by the black hole presence only by means of the halo solution:
\begin{eqnarray}
  \rho_{h} \equiv \phi ~ \phi^*
= \rho_s\left|{}_{1}F_1 \left(1-i\alpha m r_s,2,i2x\right)\right|^2 ,
\label{eq:rho-total}
\end{eqnarray}
where $\rho_s$ is a parameter and $_{1}F_1$ corresponds to the hyper-geometric function of order $(1,1)$. We have defined the dimensionless quantities 
$\alpha=Mm$ and $x\equiv kr = r/r_s$. It is interesting to notice the order of magnitude of the parameters involved in the solution: 
for an ultra-light scalar field with $m\sim 10^{-22}eV$ we have $m^{-1}\sim 0.5 \, pc$,
and for a black hole with mass $M\sim 10^{10} M_\odot \rightarrow$ $(Mm^2)^{-1}\sim kpc$,
which has the same order of magnitude of the core size and the galactic bulge size. Taking advantage of these scaling relations, we choose the following parametrization for the characteristic size $r_s=1/k$ of the solution:
\begin{eqnarray}\label{rsgammaalpha}
r_s=\frac{\gamma}{\alpha m} ,
\end{eqnarray}
where we introduced $\gamma$ as free parameter, that scales $r_s$ in units of $(Mm^2)^{-1}$. Using this parametrization, the dimensionless density of the SF configuration reads as
\begin{eqnarray}
\hat{\rho}_{h}(x;\gamma)\equiv\frac{\rho_{h}(x;\gamma)}{\rho_{s}}=\left|{}_1 F_1 \left(1-i\gamma,2,i2x\right) \right|^2 ,
\end{eqnarray}
that we show in Fig.~\ref{figrhodm} for different values of $\gamma$.

\begin{minipage}{0.43\textwidth}
\vspace{5mm}
\centering
  \includegraphics[width=\textwidth]{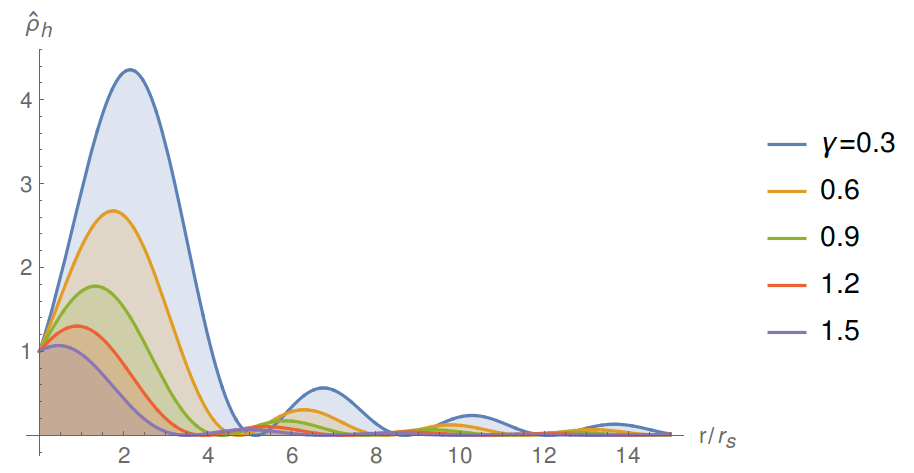}
\captionof{figure}{\small{Dimensionless radial density profile for the SF configuration for different values of the parameter $\gamma \equiv m^2 M / k$. The configuration takes its peak closer to the galactic center as its size increases for a fixed value of the black hole mass $M$.}}
\label{figrhodm}
\vspace{5mm}
\end{minipage}
\newline

In this way, we define conveniently the dimensionless DM density profile as function of the dimensionless variable $x$, parametrized with a single free parameter $\gamma$. From integrating out the density profile within a given radius, we obtain the radial mass density profile for the halo as
\begin{eqnarray}
  M_{h}(r) = \int_0^{r}  \rho_{h}(r) r^2 dr = r_s^3 \rho_s \hat{M}_h(x) ,
\end{eqnarray}
where we define the dimensionless mass profile as
\begin{eqnarray}
  \hat{M}_{h}(x) \equiv \int_0^{r/r_s} \hat{\rho}_{h}(x) x^2 dx .
\label{eq:hat-M}
\end{eqnarray}
%

  At this point, the only information we have about the free parameters of these profiles, $r_s$ and $\rho_s$, (or alternatively $\rho_s$ and $\gamma$) 
  has been derived from
 the condition of the existence of bound-solutions in the KG equation determined by the structure of the effective potential~\eqref{eq:Veff}. As a consequence,
 a lower bound for $r_s$ is established and by boundary conditions we know that \citep{lee:2015}
\begin{eqnarray}\label{eq:boundary}
	\rho_s\sim \alpha .
\end{eqnarray} 
However, the remaining indetermination  
is going to be removed by using information from some universal features of galaxies.    
In the next section 
we shall impose such constraints over the central density and in section~\ref{sec:visible-matter}
we shall estimate $r_s$ from the observed 
 correlation between $M$ and the velocity dispersion of stars in galaxies.

\section{Constraints on the Central DM Density from Mass-Acceleration-Discrepancy- Relation}
\label{sec:constraints}

Recently, it has been shown from the observed rotation curves of 153 galaxies from the SPARC database \citet{Lelli:2016}, including galaxies with very different features and morphologies with high-resolution gas and stars information, that the acceleration inferred from the observations strongly correlates with the acceleration due to the baryonic matter, showing a mass-discrepancy at the value $g^\dagger = 1.2 \times 10^{-10}~m/s^2$ \citep{McGaugh:2016PRL,McGaugh:2017}. Such relation can be interpreted as a correlation between the baryonic and the dark matter, and moreover, the maximum radial acceleration purely produced by baryonic matter and that of dark matter are closely related, and in case that DM particles exist, the maximum radial acceleration they can reach in all halos, $g_h$, cannot be greater than an universal value given by \citep{Urena-Robles-Matos:2017}
\begin{eqnarray}\label{gmaxobs}
	g_h^{max}=0.65 g^\dagger = 7.15\times 10^{-11}m/s^2 .
\end{eqnarray}

On the other hand, the acceleration profile comes as a theoretical prediction for every model and its maximum value should be restricted by the last value. In our case, specific values of the parameter $\rho_s$ for the SF density profile~\eqref{eq:rho-total} (given a mass $M$ of the BH) are required in order to predict the value~\eqref{gmaxobs}. The modulus of the radial acceleration profile for the bosons forming the halo can be computed in terms of the mass profile as follows 
\begin{eqnarray}\label{eq:DMacceler1}
	g_h(r;\gamma,\alpha) =  \frac{G M_h(r;\gamma,\alpha)}{r^2} ,
\end{eqnarray}
which in terms of the dimensionless mass profile~\eqref{eq:hat-M} and the dimensionless independent variable $x=r/r_s$ can be easily rewritten as
\begin{eqnarray}\nonumber
	g_h(r;\gamma,\alpha)&=&   G \, r_s\rho_s \, \frac{M_{dm}(x)}{x^2},\\
	&=& G \, \mu_{dm} \, \hat{g}_h(x;\gamma) ,
\label{eq:DMacceler2}
\end{eqnarray}
where we have defined the dimensionless acceleration profile as 
\begin{eqnarray}
	\hat{g}_h(x;\gamma)\equiv\frac{\hat{M}_h(x;\gamma)}{x^2} ,
\end{eqnarray}
and we used the surface density definition given in \citet{Urena-Robles-Matos:2017}:
\begin{eqnarray}
	\mu_{dm}\equiv r_s\rho_s.
\end{eqnarray}

By imposing constraint~\eqref{gmaxobs} over the theoretical maximum of the acceleration profile, we obtain a value for the halo's central surface 
density given by 
\begin{eqnarray}\label{eq:mudm}
 \mu_{dm}=\frac{0.65 g^\dagger}{G ~ \hat{g}_h^{max}(\gamma)} ,
\end{eqnarray}
and consequently the central value for the density profile
\begin{eqnarray}\label{eq:rho0alpha}
 \rho_{s}= \alpha\frac{\rho^\dagger}{\gamma~\hat{g}^{max}_{h}(\gamma)}.
\end{eqnarray}
where 
\begin{eqnarray}\label{eq:rhodagga}
 \rho^\dagger= \frac{0.65~m~g^\dagger}{G}= 0.74 \times 10^{3}\frac{M\odot}{pc^3}.
\end{eqnarray}
 Notice that \eqref{eq:rho0alpha} is consistent with the boundary condition \eqref{eq:boundary}.
\begin{minipage}{0.4\textwidth}
\vspace{5mm}
\centering
\includegraphics[width=\textwidth]{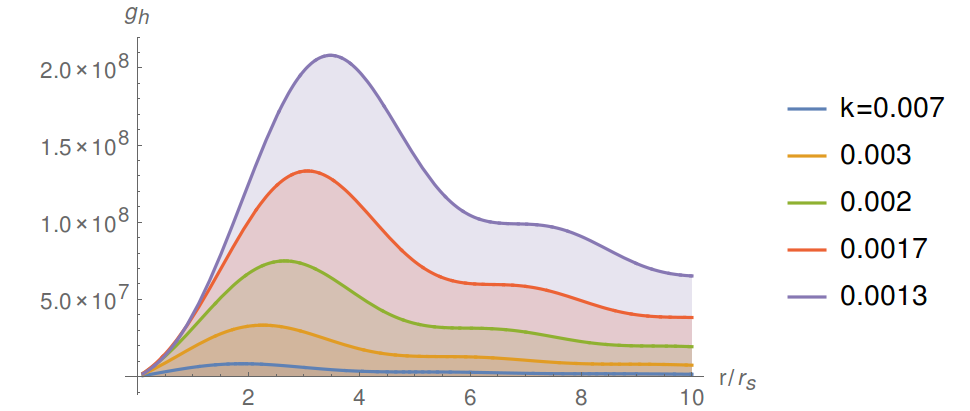}
\captionof{figure}{\small{For a fixed mass of the black hole $M\sim 10^{10} M_\odot$, the radial acceleration profile $g_h$ is plotted for different wave-lengths of the solution. Larger profiles have larger maximum accelerations.}}
\label{fig:gvsk}
\vspace{5mm}
\end{minipage}
In \citet{Urena-Robles-Matos:2017}, using the universal maximum acceleration value for DM profiles solutions of the Schroedinger-Poisson system (valid at the Newtonian limit only), they conclude that the central surface density $\mu_{dm}\equiv \rho_s r_s$ is an universal constant. In the context of the ultra-light SFDM model, $\mu_{dm}=648~M_{\odot}pc^{-2}$. This result brings as a consequence that the so-called Wave-DM soliton profile should be an universal feature of the DM halos. However, in the context of this work, where central black holes in galaxies manifestly affect the DM profile in the core and also the central density, while it is true that $g_{h}^{max}$ is a constant, from equation~\eqref{eq:mudm} it is clear that the central density $\mu_{dm}$ is not necessarily constant. 
Rather, it obeys the following  implicit relation between $\rho_s$ and $r_s$:
\begin{eqnarray}\label{eq:mudmconst}
	r_s \rho_s - \mu_{dm}(\gamma \equiv Mm^2\,r_s) = 0 .
\end{eqnarray}
Therefore the universality of the WDM soliton profile  
for galaxies hosting a SMBH not necessarily holds unless $Mm^2\,r_s= constant$.
Two interesting points arise from this last conclusion: 
Firstly, halos of galaxies with not-too-massive black holes in their centers nearly satisfy $Mm^2\,r_s=1$. This can be realized using the analysis made in subsection~\ref{sec-spectrum}; there we saw that for small masses $k\rightarrow 0$, in which case the SF profile can be described by the Ure\~na-Liddle solution \citep{UrenaLiddle:2002du} for which $r_s$ can be read off as $(Mm^2)^{-1}$.
Secondly, assuming the universality of the maximum acceleration~\eqref{gmaxobs} actually holds, in the case of observing galaxies that violate the universal WDM soliton profile, variations to the constraint~\eqref{eq:mudmconst} can lead us to consider the existence of a BH in the center of such galaxies. Consider we were studying the 
features of a SDFM halo for an specific galaxy and from fits of its rotation curve, or other set of observations, it turns out that 
$\mu_{dm}$ differs from the universal value proposed in \citet{Urena-Robles-Matos:2017}; therefore, we could set as a possibility the 
existence of a BH in its center, and from the inferred value of $\mu_{dm}$ the mass of the black hole can be estimated.

In order to completely fix the central density of our SFDM profiles we still need to fix $r_s$ (or $\gamma$). In the next section we shall fix $r_s$ by fitting the observed correlation between the mass of the black hole $M$ and the velocity dispersion of baryonic matter inside the stellar bulge \citep[see e.g.][]{Ferrarese2000}. However, at this point, we are able to set some restrictions over $\rho_s$ from theoretical grounds. The following subsection is devoted to that purpose.

\subsection{THEORETICAL UPPER BOUND FOR THE CENTRAL DENSITY OF THE SCALAR FIELD CONFIGURATION}
\label{sec:upper-bound}

  In section~\ref{sec-spectrum} we derived a cut-off for wave-number of the SKGE solutions for a range of values of the black hole mass. Correspondingly, here we derive the central density from the maximum acceleration universality  by using the procedure explained at the beginning of this section. In the following figure, the correlation between the mass of the BH and its corresponding central density is plotted. 
  
  \begin{minipage}{0.4\textwidth}
\vspace{5mm}
\centering
 \includegraphics[width=\textwidth]{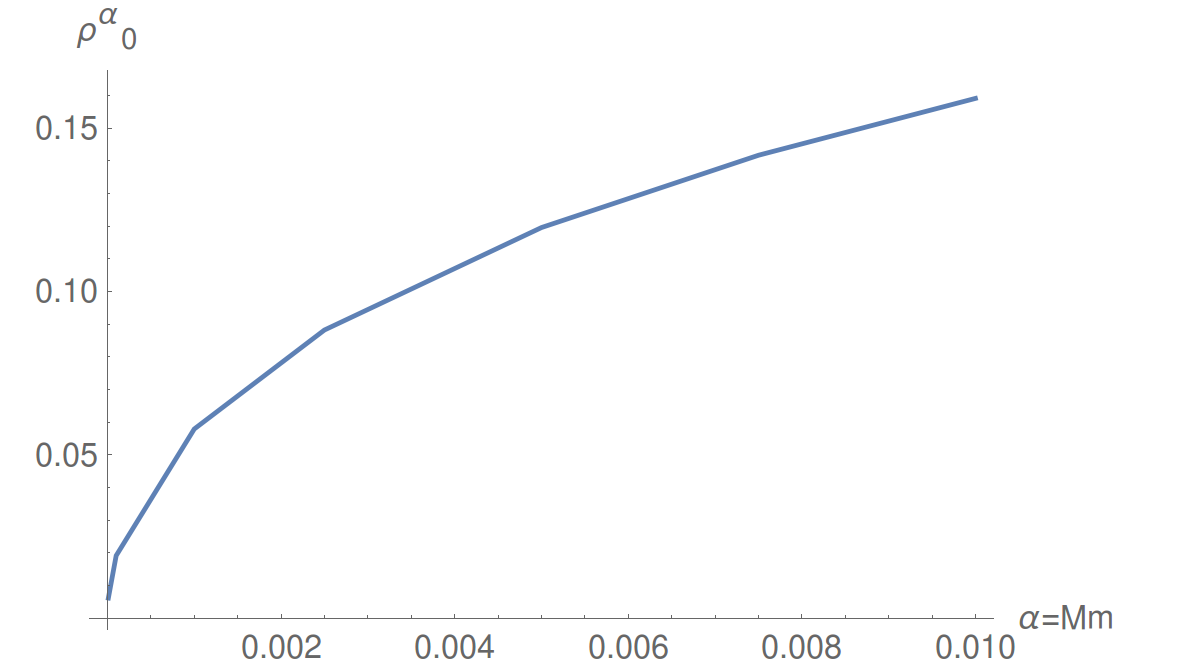}
 \includegraphics[width=\textwidth]{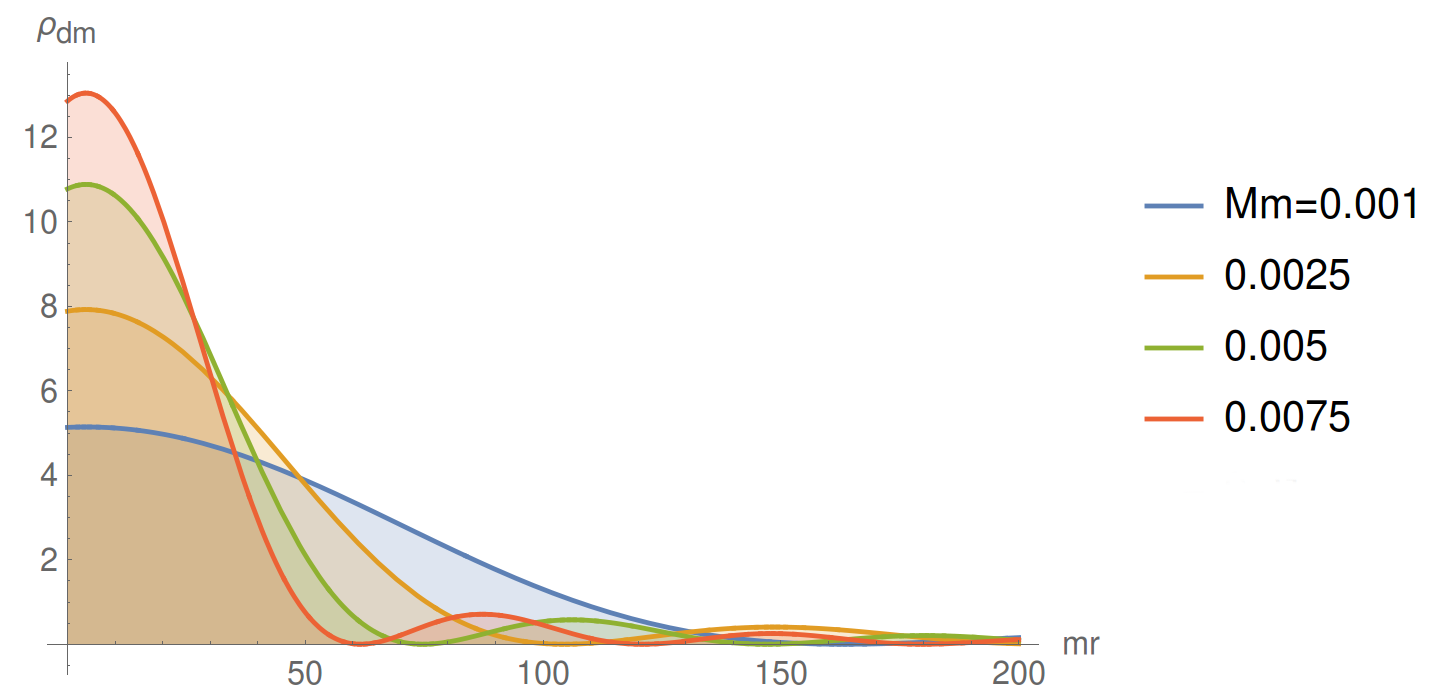}
\captionof{figure}{\small{Top: The central density value, ${\rho^\alpha}_0$, as function of the mass of the BH assuming that the scalar field profile has $r_s=1/k_{max}$. Bottom: Density profiles for different values of $\alpha=Mm$ and $r_s^{min}$. This values of the central density are obtained from the universality of the maximum acceleration constraint. As expected, the halo is denser at the center for more massive black holes; effectively, the BH increases the height and shrinks the length of the SF profile. This is in agreement with our toy model solution discussed in section~\ref{sec:driving}.}}
\label{fig:rhoalpha}
\vspace{5mm}
\end{minipage}

  According to this relation, the intuitive and ideal picture drawn in section~\ref{sec:driving} comes to be confirmed: the larger the mass of the central black hole is, the denser is the surrounding scalar field configuration. This would be actually true if nature would choose $r_s=1/k_{max}$; however, the complexity of baryons in real galaxies is mostly likely to mess up such assumption. Nonetheless, it gives us an idea of how the halos would ideally be without baryons and also sets up some restrictions over the central density of course.      
 For a fixed value of the mass of the black hole, $M$, the theoretical radial acceleration strongly depends on $r_s$. As Fig.~\ref{fig:gvsk} shows, as $k$ increases $g_h^{max}$ decreases, and together with equation~\eqref{eq:rho0alpha}, $\rho_s(\alpha,k_{max})$ increases. Therefore, for a fixed $M$, the value $\rho_s(\alpha,r_s^{min})$ 
 is an upper bound for all its possible values.  


\section{Kinematics of visible matter inside the gravitational potential of the SFDM halo}
\label{sec:visible-matter}

In practice, the mass of a black hole is derived from measurements of different features of the velocity field of stars and gas moving along the well-potential produced by the whole system including the halo, stars, gas and the black hole. From theoretical grounds, the procedure goes the other way round: we assume a black hole fully specified by its mass and we derive some kinematic features of the velocity field of the system.

In this section we finally determine the free parameters of our theory; we do this by fitting the observed correlation between the mass of the black hole and the velocity dispersion of stars and gas derived in \citet{McConnell:2012}. In order to do so, we construct a description of the complete galactic system presented in section~\ref{sec:profiles}. We assume that the visible part of the galaxy has elliptical morphology, we choose this since observations suggest that SMBHs live in massive and dispersive galaxies, which in many cases have this shape.

We have already fixed the central density and  
set $k_{max}$, however that is not enough to figure out whether or not this model actually is able to describe nature. With a view to address this problem and to find a law of behavior of $r_s$, later in subsection~\ref{sec:dmdominates}, we shall consider the ideal case in which the DM contribution of the gravitational well-potential dominates such that the velocity dispersion of stars only depends on the properties of DM and the ratio of $r_s$ and the effective radius of the galaxy. By reproducing the phenomenological $M-\sigma$ relation reported in \citet{McConnell:2012}, we construct a $r_s-M$ correlation for this ideal case. Besides, based on this determination of $r_s$, we obtain $\mu_{dm}$ as function of $\gamma$, which allows us to generalize the result of \citet{Urena-Robles-Matos:2017} in the case of galaxies hosting SMBHs.
 Finally, in subsection~\ref{sec:realgalax} we constrain the hypothetical characteristic size of the SSFDM configuration, $r_s$, corresponding to real cases by fitting the velocity dispersions of some real specific galaxies and galactic cores.
As usual, we computed the velocity dispersion by solving the Jeans equation, derived from the spherically symmetric non-collisional Boltzmann equation governing the distribution of stars and gas in the galaxy.

\subsection{VISIBLE MATTER IN THE GALAXY AND THE JEANS EQUATION}
\label{sec:DensMass}

 Because we are assuming spherical symmetry, our model predicts that stars only move along the radial direction. The stellar spatial distribution is fully described by the distribution function $f(r,v)$, which is the probability of finding a star at radius $r$ with velocity $v$. This distribution
 satisfies the Boltzmann equation and once it is known, all the macroscopic statistical quantities associated to the visible part of the galaxy can be determined. 
However, to determine this distribution is not straightforward and sometimes it is not even necessary to compute some observables, as it happens with the dispersion of velocities which obeys the Jeans equation. This relation can be derived from the former in the case that stars in the system are non-collisional and when spherical symmetry is imposed and is given by 
\citep{binney-tremaine}:
\begin{eqnarray}\label{eq:Jeans}
\frac{1}{\rho_*(r)}\frac{d(\rho_* \sigma_{*}^2)}{dr}+\frac{2\beta\sigma_{*}^2}{r}=-\frac{GM_{tot}(r)}{r^2},
\end{eqnarray}
where $\beta$ is the anisotropy parameter which we shall ignore since, even in the more complex cases, it modifies the results no more than $5\%$ \citet{binney-tremaine,McConnell:2012}; the total mass of the galactic system, $M_{tot}$, and the stellar velocity dispersion, $\sigma_*$, are defined by
\begin{eqnarray}
M_{tot}(r)&=&M_{bar}(r)+M_{dm}(r),
\end{eqnarray}
with $\bar{v}$ the mean radial velocity and $M_{bar}$ and  $M_{dm}$ the baryonic mass (gas and stars) and dark matter, respectively, enclosed inside a given radius. In this work we assume $M_{bar}$ can be described by the Plummer mass profile (see equation~\eqref{eq:Plummer}) which is typically used to describe stars in galactic bulges where the velocity dispersion is normally observed.

The left side of equation~\eqref{eq:Jeans} corresponds to kinematic terms of the visible matter, while the right part involves the dynamical sources that produce the galactic well-potential that triggers the kinematics. 
The radial density distribution of stars  
is commonly described by the Plummer density profile given by   
\begin{equation}
	\rho^*(r) = \frac{\rho^*_0} {\left[ 1 +(r / R_{eff})^2 \right]^{5/2}} ,
\label{eq:Plummer}
\end{equation}
where $\rho^*_0\equiv 3M_{tot}^*/(4\pi R_{eff}^3)$ is the stellar central density and $M_{tot}^*$ is the mass of stars enclosed within the effective radius $R_{eff}$, which is defined as the radius at which the luminosity of the galaxy decreases to a half of its central value.

We define the dimensionless density profile $\hat{\rho}^*\equiv\rho^*/\rho^*_0$ as function of the tilded variable $\tilde{r}\equiv r/R_{eff}$, illustrated in Fig.~\ref{fig:rhoPlummer}.

\begin{minipage}{0.4\textwidth}
\vspace{5mm}
\centering
\includegraphics[width=\textwidth]{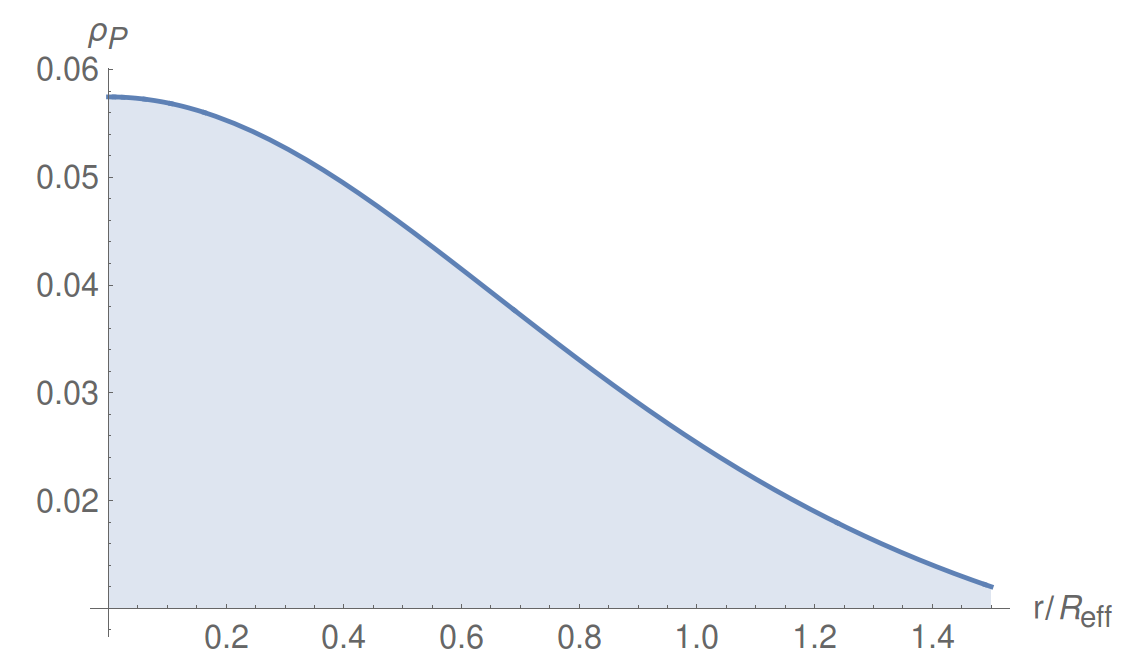}
\captionof{figure}{\small{Density profile of stars and gas according to the Plummer model, suitable for describing bulges. Notice that this is a core-like distribution which is compatible with SFDM models.}}
\label{fig:rhoPlummer}
\vspace{5mm}
\end{minipage}

The corresponding Plummer mass profile is obtained by integrating the density profile~\eqref{eq:Plummer} within a radius $r$. The dimensionless Plummer mass profile of stars, $\hat{M}^*$, is defined a
%
\begin{eqnarray}\nonumber
  \hat{M}^*(r) \equiv \frac{M^*(r)}{M_{tot}^*} =  \frac{\left(r/R_{eff}\right)^3}{\left[ 1 
  + \left(r/R_{eff} \right)^2 \right]^{3/2}} .
\end{eqnarray}

\subsection{CORRELATION $M-r_s$ IN DARK-MATTER DOMINATED GALAXIES FROM THE OBSERVED $M-\sigma^*$ RELATION}
\label{sec:dmdominates}

The mass of visible matter in galaxies is an important parameter to consider in order to understand the co-evolution of the black hole, the dark matter halo and the messy system of stars and gas. Real galaxies hosting SMBHs usually contain an important fraction of visible matter in-falling into the gravitational well potential and thanks to such component these objects can be detected. In the following subsection we consider that case for very specific situations. However in this part, we would like to focus in studying the influence of the structure of the halo into the velocity dispersion of a subdominant fraction of visible matter. For this study, let us consider that the gravitational potential well is mainly produced by the scalar field configuration forming the dark matter halo. The Jeans equation corresponding to such situation can be written as  
\begin{eqnarray}\label{eq:JeansEq}
\frac{1}{\hat{\rho}_*(r)}\frac{d(\hat{\rho}_*\sigma_{*}^2)}{dr}&=& -\frac{GM_{dm}(r)}{r^2}.
\end{eqnarray}
Notice that $M_{tot}^*$ is not a free parameter anymore. Equation~\eqref{eq:JeansEq} is equivalent to 
\begin{eqnarray}\label{eq:IntJeansEq}
	\sigma_{*}^2(r) = \frac{1}{\hat{\rho}_*(r)}\int_0^r\frac{GM_{dm}(r')~\hat{\rho}_*(r')}{r'^2}\,dr',
\end{eqnarray}
which written in terms of dimensionless quantities 
turns into 
\begin{eqnarray}\nonumber
	\sigma_{*}^2(x) = \frac{G\,\rho^\dagger}{m^2\alpha}\frac{\gamma}{\hat{g}_{dm}^{max}(\gamma)}\frac{1}{\hat{\rho}_*(x,a)}\int_0^{x}\frac{G\hat{M}_{dm}(x',\gamma)~\hat{\rho}_*(x',a)}{x'^2}\,dx' , \\
\label{eq:sigma2dimensionless}
\end{eqnarray}
where we have redefined the Plummer profile in terms of the dark-to-visible-size ratio $a \equiv r_s/R_\mathbf{eff}$ as
\begin{eqnarray}
\hat{\rho}_*(x,a)\equiv \frac{1}{\left(1 + a^2x^2\right)^{5/2}}.
\end{eqnarray}

It is worth to notice that equation~\eqref{eq:sigma2dimensionless} results to be a product of a term depending only on $\gamma$ and other only depending on $\alpha$. 
The square of the stellar velocity dispersion evaluated in the effective radius is given by 
\begin{eqnarray}\nonumber
	\sigma_{*}^2(R_{eff}) = \frac{(v^\dagger)^2}{\alpha}~\sigma_{(a)}^2(\gamma)
\end{eqnarray}
where we have introduced the scalar field characteristic velocity $v^\dagger$ and the dimensionless quantity $\sigma_{(a)}^2(\gamma)$:
\begin{eqnarray}\nonumber
	\sigma_{(a)}^2(\gamma)&\equiv&\frac{\gamma}{\hat{g}_{dm}^{max}(\gamma)}\frac{1}{\hat{\rho}_*(1)}\int_0^{1/a}\frac{G\hat{M}_{dm}(x',\gamma)~\hat{\rho}_*(x',a)}{x'^2}\,dx' ,\\
\label{sigmaagamma}\\
(v^\dagger)^2&\equiv&\frac{0.65~g^\dagger}{m}=1.108\times 10^6 ~ m^2 s^{-2}.
\label{eq:vdagger}
\end{eqnarray}

Notice that the square of the actual velocity dispersion is just a rescaling of~\eqref{sigmaagamma} which exclusively depends on $\alpha$ and the mass of the scalar field, $m$.
The last point is interesting because, unlike other observables, the dependence on $\alpha$ and $m$ in  $\sigma_{(a)}^2$ are separated and hence they are not degenerated. 
In Fig.~\ref{fig:sigmaagamma}, $\sigma_{(a)}^2(\gamma)$ is plotted for some values of the dark-to-visible-size ratio $a$, for a fixed mass $M$ of the black hole; notice that
as the size of the halo increases in relation to the size of the bulge, the whole velocity dispersion profile is suppressed.
By observing the integral in~\eqref{sigmaagamma}, it can be noticed that the stellar density profile serves as a weight of the radial acceleration profile, 
therefore, the larger $a$ is, such weight becomes steeper and the integrand falls down at smaller radius. This suggest a connection between baryons and their
hosting SSFDM halos, that is, for a fixed $M$ and $R_{eff}$,
visible matter in galaxies is less dispersive if it is embedded in larger halos. 

\begin{minipage}{0.44\textwidth}
\vspace{5mm}
\centering
 \includegraphics[width=\textwidth]{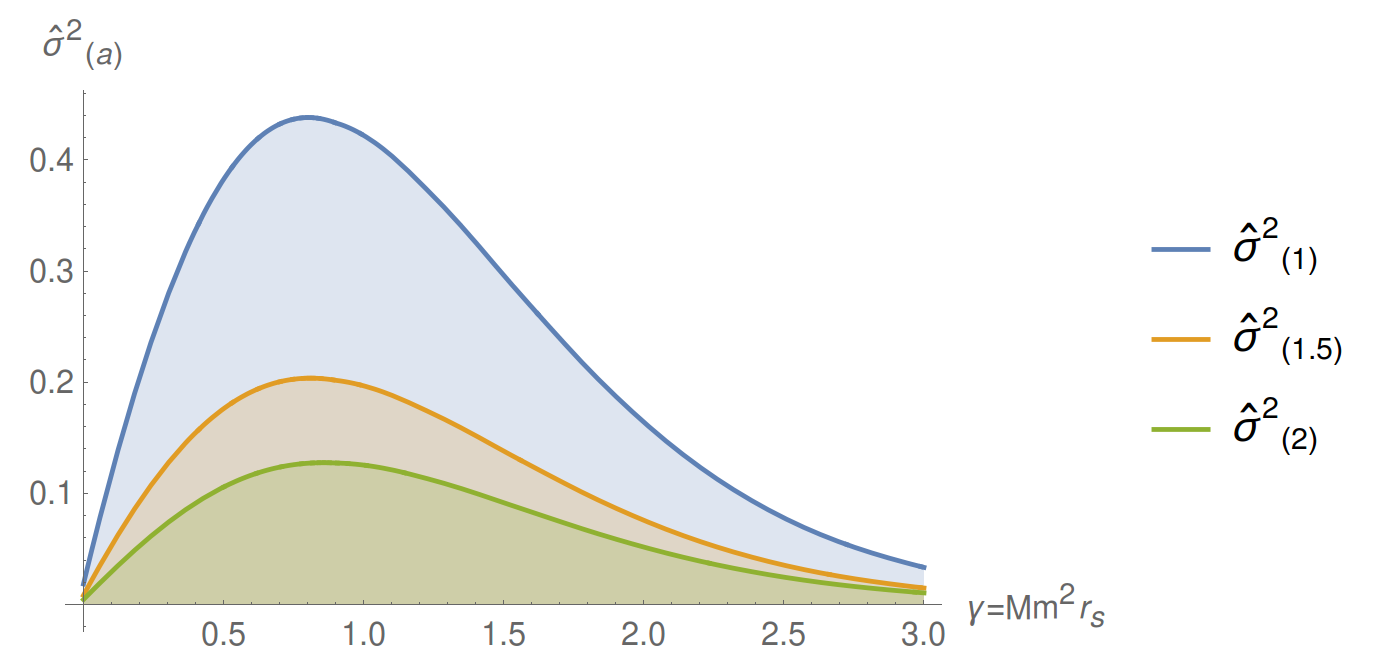}
\captionof{figure}{\small{Dimensionless velocity dispersion $\sigma_{(a)}^2$ as function of $\gamma$. If the size of the halo increases in relation to the size of the bulge, 
the velocity dispersion profile is suppressed.}}
\label{fig:sigmaagamma}
\vspace{5mm}
\end{minipage}

Let us turn to determine the values $r_s$ for a fixed $M$ by fitting the phenomenological $M-\sigma$ relation obtained by \citet{McConnell:2012} for a large 
sample of galaxies of different morphologies, time of formation and masses with the common feature of having a large stellar and gas velocity dispersion. 
The best fit for the set of $M-\sigma$ points of the sample is given by  
\begin{eqnarray}\label{eq:phenomMsigma}
\log_{10} \left( \frac{M}{M_\odot} \right) =8.32 + 5.64 \log_{10} \left( \frac{\sigma}{200 \ km\,s^{-1}} \right) .
\end{eqnarray}

\begin{minipage}{0.44\textwidth}
\vspace{5mm}
\centering
 \includegraphics[width=\textwidth]{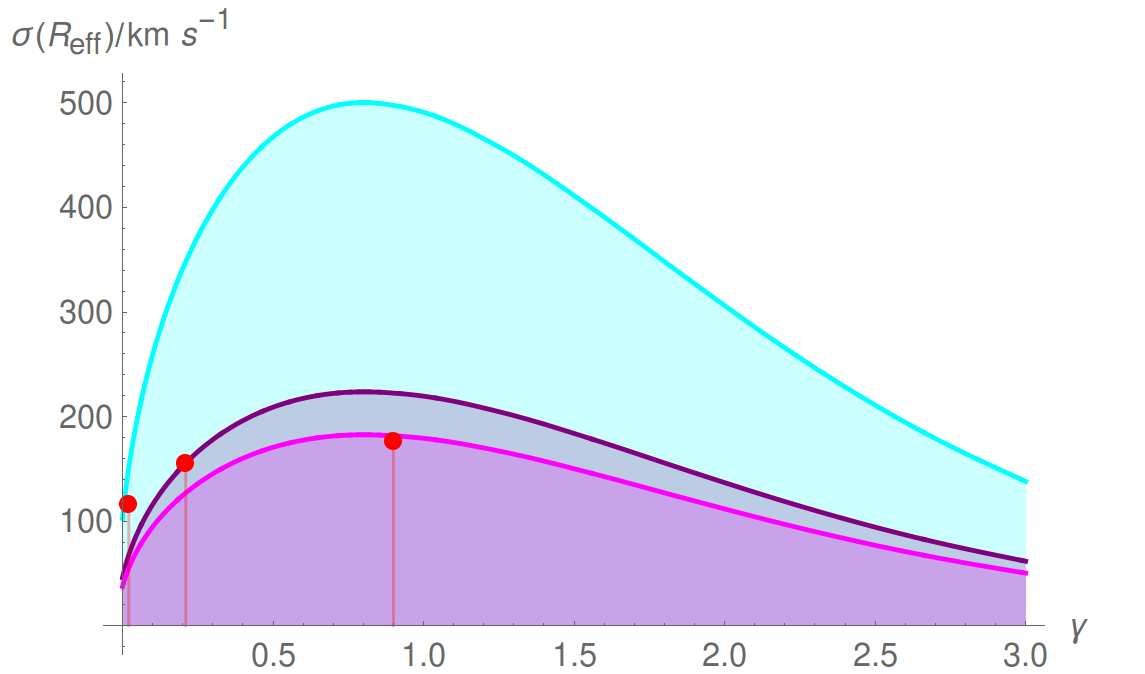}
\captionof{figure}{\small{Lines correspond to theoretical $\gamma-\sigma$ curves for $M=10^7M_\odot$(cyan), $M=5\times 10^7M_\odot$(purple)  and $M=8\times 10^7M_\odot$ (magenta) 
Red dots correspond to picked values of $\gamma$ that satisfy the phenomenological $M-\sigma$ relation~\eqref{eq:phenomMsigma}. }}
\label{fig:sigmagamma}
\vspace{5mm}
\end{minipage}

We generated a set of bins of values of $M$ running from $10^6$ to $10^9$ and for each bin we used the prescription explained above to compute $\sigma$. By 
trying different values of $\gamma$, we picked the one that reproduced the value of $\sigma$ associated to $M$ by relation~\eqref{eq:phenomMsigma}. Fig.~\ref{fig:sigmagamma} shows theoretical $\sigma-\gamma$ curves corresponding to different masses of the black hole and the picked values reproducing the observational 
value of $\sigma$. We should mention here that the model fails to reproduce $\sigma$ for masses $M$ larger than $10^9 M_\odot$.
The result of this procedure is summarized in Fig.~\ref{fig:rsM}.

\begin{minipage}{0.44\textwidth}
\vspace{5mm}
\centering
 \includegraphics[width=\textwidth]{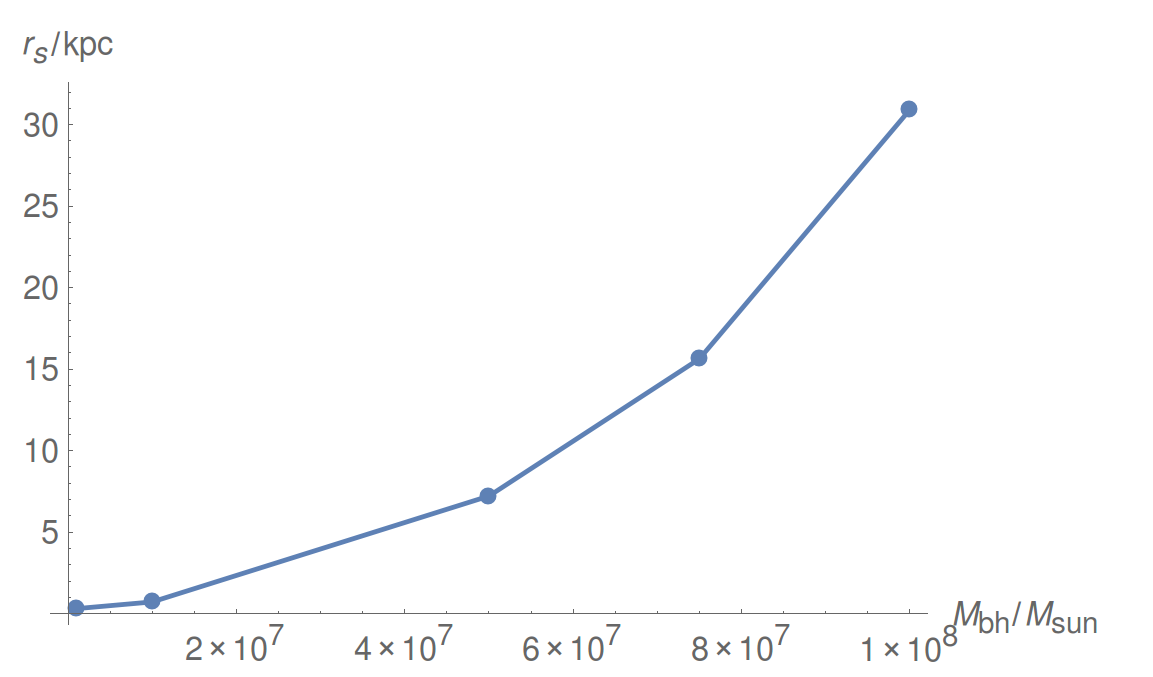}
\captionof{figure}{\small{Scaling relation between the characteristic length of the SSFDM halo and the mass of the hosted SMBH for dark-matter dominated bulges. This relation is derived from fitting the observed $M-\sigma^*$ relation. }}
\label{fig:rsM}
\vspace{5mm}
\end{minipage}

A complementary result of this work is the corresponding constraint~\eqref{eq:mudmconst} for these values of $r_s$. As mentioned before, unlike the 
standard SFDM profiles, in the case of SSFDM profiles $\mu_{dm}$ is not a constant, rather, as Fig.~\ref{fig:mudm} shows, it depends on the parameter
$\gamma = Mm^2r_s$. It is interesting to notice that for values tending to unity, $\mu_{dm}$ tends to a constant.

 \begin{minipage}{0.44\textwidth}
\vspace{5mm}
\centering
 \includegraphics[width=\textwidth]{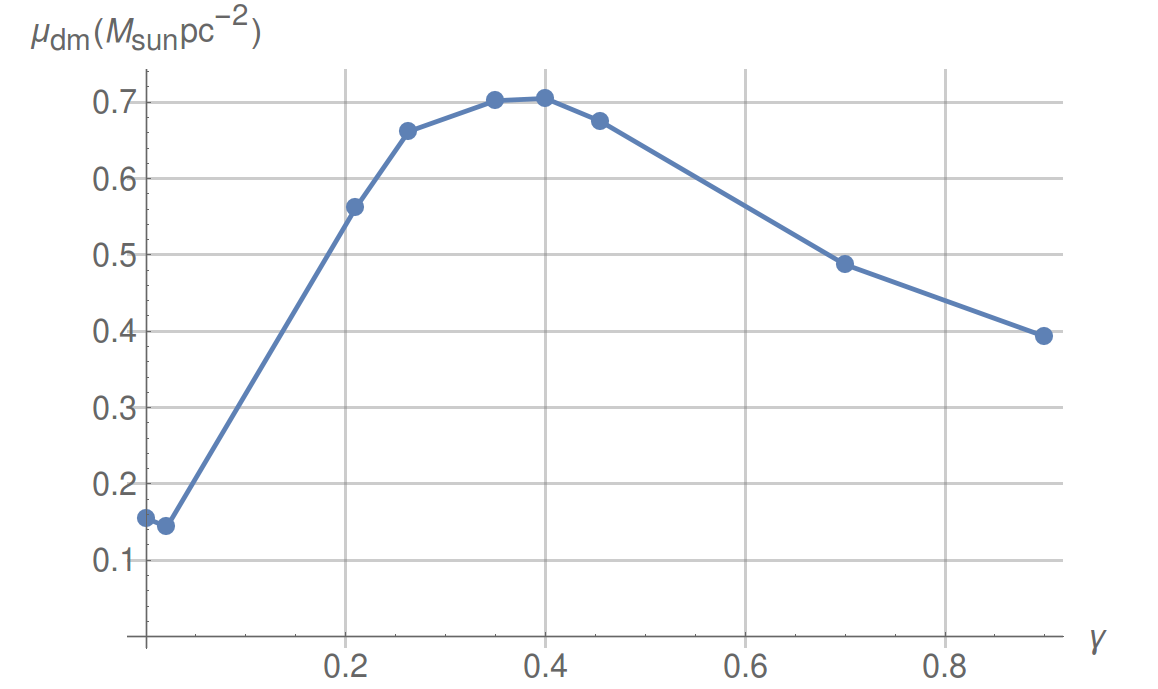}
  \captionof{figure}{\small{Surface 
  density for several values of $\gamma$. As expected in this model, this quantity is not a constant, however for large values of 
  $r_s$ it tends to a constant asymptotically.}}
\label{fig:mudm}
\vspace{5mm}
\end{minipage}

\subsection{CORRELATION $M-r_s$ FOR REAL GALAXIES}
 \label{sec:realgalax}
 
We solved numerically equation~\eqref{eq:JeansEq} using the Plummer profile for the visible part and the mass profile corresponding to SSFDM described before for different elliptic galaxies hosting SMBH in their centers. The corresponding parameters for each galaxy are enlisted in Table~\ref{table-galaxies}; they were extracted from catalogs in \citet{McConnell:2012}. \citet{Larkin:2016} describe a correlation between the stellar total mass and the effective radius and they infer the mean-stream trend by means of the following phenomenological relation:
\begin{eqnarray}
R_{eff}/kpc =  1.5*\bar{M}^{0.1} (1 + \bar{M}^5)^{0.1},
\end{eqnarray}
where $\bar{M} \equiv M_{tot}^*/2\times 10^{10}M_\odot$.
 We used this result in order to estimate the effective radius of some galaxies in our catalog (Table~\ref{table-galaxies}) labeled with $*$. 

Table~\ref{table-galaxies} summarizes the parameters used to solve the Jeans equation for each case. 

\begin{minipage}{0.4\textwidth}
\vspace{5mm}
\centering
\begin{tiny}
\begin{tabular}{ccccc}
\hline
Galaxy&$M_\mathrm{BH}/M_\odot$ & $\alpha$ & $M_{tot}^*/M_\odot$ & $R_{eff}/kpc$ \\ 
\hline \hline
Milky Way Bulge&$4.1\times 10^{6}$ & $7.954\times10^{-7}$ & $9\times 10^{9}$ & $1.4$\\
N3384&$1.1\times 10^{7}$ & $2.13\times10^{-6}$ & $1.9\times 10^{10}$ & $1.58^*$\\
N3585&$3.2\times 10^{8}$ & $6.4\times10^{-5}$ & $1.6\times 10^{11}$ & $5.2^*$\\
N3608&$4.7\times 10^{8}$ & $9.1\times10^{-5}$ & $7.66\times 10^{10}$ & $3.36^*$\\
N3379&$4.2\times 10^{8}$ & $8.2\times10^{-5}$ & $6.86\times 10^{10}$ & $3.143^*$\\
M87&$3.2\times 10^{9}$ & $6.208\times10^{-4}$ & $3.3\times 10^{11}$ & $8.0$\\
M49 & $2.5\times 10^9$ & $4.85 \times 10^{-4}$ & $4.2 \times 10^{11}$ & 9.3 \\
NGC 4889 & $2.1 \times 10^{10}$ & $4.074 \times 10^{-3}$ & $9.5 \times 10^{11}$ & 15.2 \\
\hline
\end{tabular}
\end{tiny}
\captionof{table}{\small{Physical parameters for SMBHs observed in several galaxies. The columns represent: the name of the galaxy, the inferred SMBH mass, the corresponding value of the parameter $\alpha=Mm$ for $m\sim10^{-22}~eV$, the total observational stellar mass and the effective radius.}}
\label{table-galaxies}
\vspace{5mm}
\end{minipage}

Let us take as an example the Milky Way bulge; in such case, according to our prescription, in order to reproduce the $M-\sigma$ relation, the visible matter and the dark matter profiles have nearly the same sizes, however the dark matter central density is clearly dominant in about $80\%$ over the former one (see Fig.~\ref{fig:densityMW}). Naively, this may tell us that the treatment presented in the previous subsection is a good approximation for computing the velocity dispersion of stars and gas. However, the fitting curve of $M-\sigma$ derived from ignoring the contribution of baryons to the potential well fails to fit the $M-\sigma$ point of real galaxies, as Fig.~\ref{fig:rsMbhtheovsobs} shows. Clearly, this means that the contribution of baryons comes to be important after all. However, a better fit is achieved if the pure dark matter fitting-curve is shifted by $100~km s^{-1}$. Therefore, that gap is interpreted as the velocity dispersion due to baryons.

 \begin{minipage}{0.4\textwidth}
\vspace{5mm}
\centering
 \includegraphics[width=\textwidth]{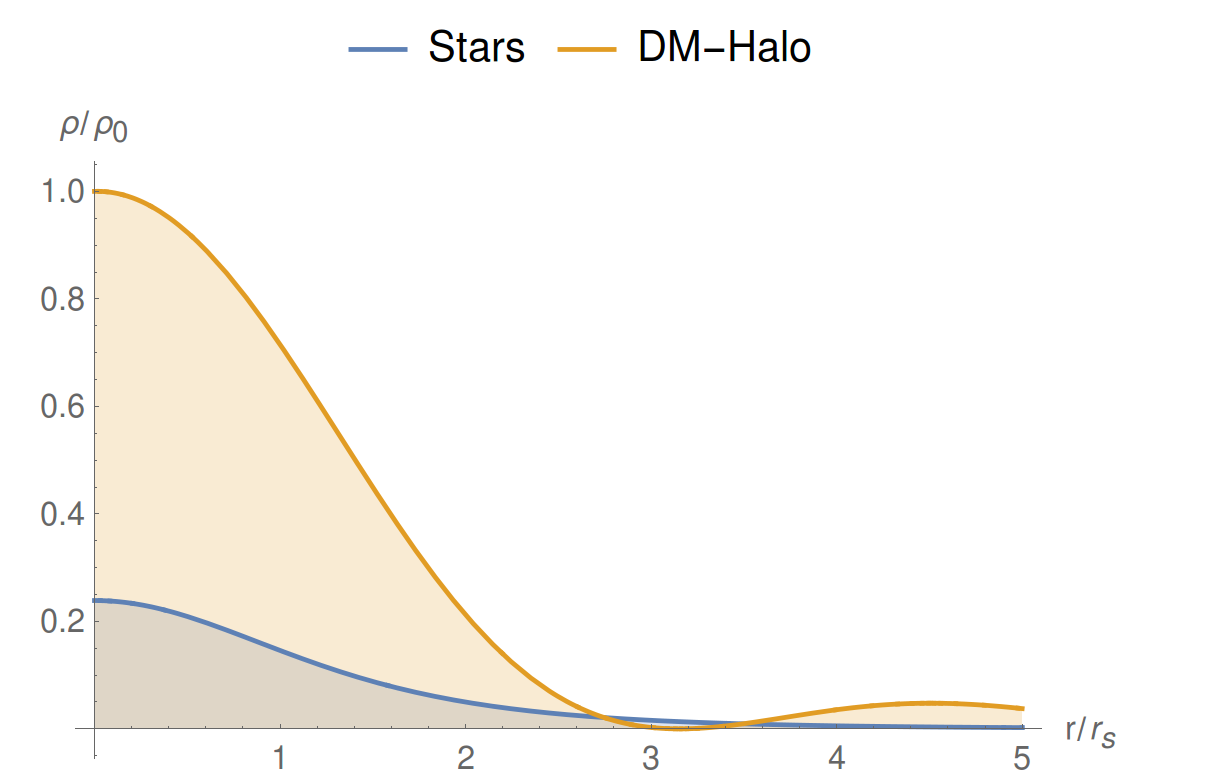}
\captionof{figure}{\small{Radial densities for both baryonic matter (Plummer profile) and the SSFDM halo for the case of the bulge of the Milky Way. Clearly, dark matter dominates, as expected.}}
\label{fig:densityMW}
\vspace{5mm}
\end{minipage}

\begin{minipage}{0.4\textwidth}
\vspace{5mm}
\centering
\begin{scriptsize}
\begin{tabular}{cccc}
\hline
Galaxy & $r_s^t/kpc$ & $\sigma(R_{eff})^t/km~s^{-1}$ & $\sigma_{lit}(R_{eff})$ \\
\hline \hline
Milky Way Bulge & 0.615 & 90.05 & 103$^a$ \\
& & & 90$^c$ \\
\hline
M87 & 3.04 & 245.166 & 324$^a$ \\
    & & & 264$^b$ \\
    & & & 245$^c$ \\
\hline
M49 & 3.44 & 264.94 & 300$^a$ \\
    & & & 250$^b$ \\
    & & & 265$^c$ \\
\hline
NGC 4889 & & & 347$^a$ \\ 
         & & & 345$^c$ \\
\hline
N3585 & 2.1 & & 213 $^a$ \\ 
\hline
N3384 & 2.85 & 143.34 & 143 $^a$ \\ 
\hline
N3379 & 1.08 & & 206 $^{a}$ \\ 
\hline
$^a$ \citep{McConnell:2012}\\
$^c$ \citep{Larkin:2016}\\
$^t$ From theory
\end{tabular}
\end{scriptsize}
\captionof{table}{\small{Here we present the results of the derived values of the parameter $r_s$ to obtain velocity dispersions close to the values reported in the literature for each galaxy in Table~\ref{table-galaxies}. 
}}
\label{table:sigma}
\vspace{5mm}
\end{minipage}

\begin{minipage}{0.44\textwidth}
\vspace{5mm}
\centering
 \includegraphics[width=\textwidth]{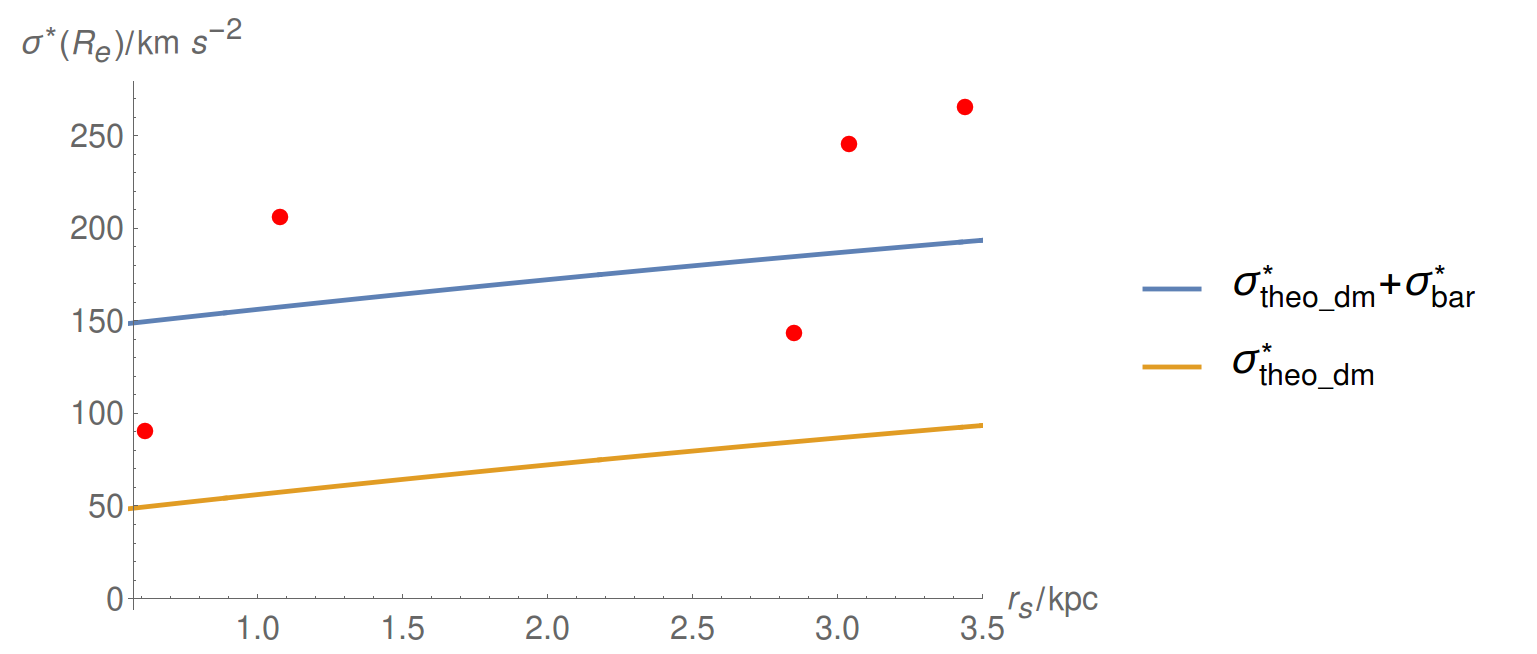}
\captionof{figure}{\small{$r_s-\sigma$ relation. The yellow line corresponds to the relation obtained from fitting equation~\eqref{eq:Msigma} in the previous subsection, that only takes into account the dark matter contribution in the well potential. Red points correspond to $r_s-\sigma$ for some galaxies in our catalog (Table~\ref{table-galaxies}) and the blue line is a rough fit curve to those points given as $\sigma_{theo_{dm}}+100~km~s^{-1}$.}}
\label{fig:rsMbhtheovsobs}
\vspace{5mm}
\end{minipage}

\section{Discussion and conclusions}
\label{sec:discussion}

In this paper, we assume the possibility that galactic systems hosting a super-massive black hole in their center were formed very early in the universe from the collapse of a Bose Einstein Condensate made of modes of a scalar field in the ground state. In our assumptions, we consider the hypothesis that the late-time dark matter halos hosting the whole system are made of quasi-resonant solutions of a real ultra-light scalar field that are being swallowed by the SMBH at such slow rate that their lifetime scales as the age of the universe. Within the most general and realistic context, the scalar field would be self-interacting and self-gravitating and, along with the metric of space-time, form a complicated system of coupled non-linear differential equations which has been studied numerically by many groups since long time ago. However, the available computing and numerical tools at the moment have allowed to explore these systems in a range of parameters corresponding to models of boson stars at most. Solutions for configurations of scalar field with the size of a galactic halo and black holes as massive as SMBH has not been obtained at the date.   
In order to turn around such technical problem, in this paper use a semi-analytical approach to model galactic systems in a quasi-static limit. As a first step towards solving the problem, in this paper we propose the simplest model to describe the structure of galactic systems hosting a super-massive black hole in its center. We model the halos of galaxies as configurations made of solutions of the Klein-Gordon equation with a Schwarzschild background. We find analytic solutions for a range of $M$ corresponding to SMBH in the limit $r>2M$, that is, when the observed is placed faraway from the black hole and inside the galactic bulge. By using such solutions we derive the corresponding density and mass profiles with $\rho_s$ and $r_s$ as free parameters, which were constrained using observational features of galaxies. In specific, $\rho_s$ was determined for each value of $M$ from the universality of maximum acceleration of dark-matter particles and later on $r_s$ was fixed by fitting measurements of the velocity dispersion of real elliptical galaxies. Additionally, we generalize the constraint $\mu_{dm}=constant$ for the central surface density of the halo for galaxies containing SMBH in their centers. By reproducing the observational points of the $M-\sigma$ relation, we derive a $r_s-M$ correlation which is a result that brings up new information about the scalar field dark matter model. 
It is worth to mention that given the recent direct observations of Sagittarius A by the EHT team \citet{RicarteDexter2015,EventHorizonTelescope}, a new era of detailed explorations of the deep-inner galactic region is about happening, this would bring up a new source of evidence of the properties of SMBH and its influence on stars laying in the galactic bulge. Particularly, observations of the stellar evolution across this region will bring wealthy information of the dark matter configurations and it will be possible to discriminate between different DM models.  Either to test our hypothesis about SMBH formation and to compare different DM models in the deep-inner galactic regions using these and upcoming direct observations of SMBH in a short future is a compelling goal which we are after.  

\section*{Acknowledgements}
This work was partially supported by CONACyT M\'exico under grants CB-2011 No. 166212, CB-2014-01 
No. 240512, Project
No. 269652 and Fronteras Project 281;
Xiuhcoatl and Abacus clusters at Cinvestav, IPN; I0101/131/07 C-234/07 of the Instituto
Avanzado de Cosmolog\'{\i}a (IAC) collaboration (http://www.iac.edu.mx).
A.A.L. and L.E.P. acknowledge financial support from CONACyT postdoctoral
fellowships.

\bibliographystyle{mnras}
\bibliography{SF-SMBH}

\bsp	
\label{lastpage}
\end{document}